\pdfoutput=1
\documentclass[12pt]{article}
\usepackage{amsmath,array,amssymb,latexsym,cite,eucal,amsthm,enumerate,multirow}
\usepackage[matrix,arrow,frame,import,curve,color]{xy}
\usepackage{dynkin-diagrams}  % includes xparse
\usepackage{tikz}
\usetikzlibrary{arrows}
\usetikzlibrary{decorations.pathmorphing}
\usetikzlibrary{decorations.markings}
\usetikzlibrary{calc}

\usepackage{graphicx}

\newtheorem{theorem}{Theorem}
\newtheorem{prop}[theorem]{Proposition}

\theoremstyle{definition}

\newif\iffigs\figstrue

\DeclareFontFamily{U}{rsf}{}
\DeclareFontShape{U}{rsf}{m}{n}{
  <5> <6> rsfs5 <7> <8> <9> rsfs7 <10-> rsfs10}{}
\DeclareMathAlphabet\Scr{U}{rsf}{m}{n}

\def\C{{\mathbb C}}

\def\R{{\mathbb R}}
\def\Z{{\mathbb Z}}
\def\H{{\mathbb H}}
\def\Int{\operatorname{Int}}

\def\la{{\langle}}
\def\ra{{\rangle}}
\def\Im{\operatorname{Im}}

\def\sl{\operatorname{\mathfrak{sl}}}

\def\SO{\operatorname{SO}}
\def\Sl{\operatorname{SL}}
\def\Gl{\operatorname{GL}}
\def\GO{\operatorname{O{}}}
\def\SU{\operatorname{SU}}
\def\GU{\operatorname{U{}}}
\def\Sp{\operatorname{Sp}}

\def\so{\operatorname{\mathfrak{so}}}
\def\su{\operatorname{\mathfrak{su}}}

\def\Span{\operatorname{Span}}

\def\frak{\mathfrak}

\newcommand{\bigslantt}[3]{{\left.\raisebox{-.2em}{$#1$}\right\backslash
    \raisebox{.2em}{$#2$}\left/\raisebox{-.2em} {$#3$}\right.}}

\def\cM{{\Scr M}}

\def\ff#1#2{{\textstyle\frac{#1}{#2}}}

\def\eqn#1#2{\begin{equation}#2
    \ifx{#1}{}\else\label{#1}\fi\end{equation}}

\newcolumntype{M}[1]{>{\hbox to #1\bgroup\hss$}l<{$\egroup}}
\makeatletter
\newcommand\@brcolwidth{0.67em}
\newenvironment{eqmatrix}{%
    \hskip-\arraycolsep
    \new@ifnextchar[\@brarray{\@brarray[\@brcolwidth]}%
}{%
    \endarray
    \hskip -\arraycolsep
}
\def\@brarray[#1]{\array{r*\c@MaxMatrixCols {M{#1}}}}
\makeatother

\begin{document}

\begin{titlepage}
\begin{flushright}
% arXiv:
July 2024%\today
\end{flushright}
\vspace{.5cm}
\begin{center}
\baselineskip=16pt
{\fontfamily{ptm}\selectfont\bfseries\huge
String Moduli Spaces and\\ Parabolic Factorizations\\[20mm]}
{\bf\large Paul S.~Aspinwall
 } \\[7mm]

{\small

Departments of Mathematics and Physics,\\ 
  Box 90320, \\ Duke University,\\ 
 Durham, NC 27708-0320 \\ \vspace{6pt}

 }

\end{center}

\begin{center}
{\bf Abstract}
\end{center}
The symmetric spaces that appear as moduli spaces in string theory and
supergravity can be decomposed with explicit metrics using parabolic
subgroups.  The resulting isometry between the original moduli space
and this decomposition can be used to find parametrizations
of the moduli. One application is to
determine the volume parameter in conformal field moduli spaces for K3
surfaces. Other applications involve simple Dynkin diagram
manipulations inducing ``going up and down'' between symmetric spaces
by adding parameters and going to limits respectively. For
supersymmetries such as $N=6$, this involves combinatorics of less familiar
``restricted'' Dynkin diagrams.

\vfil\noindent
%\verb$LastChangedRevision: 30 $\\
%\verb$LastChangedDate: 2024-07-30 10:06:44 -0400 (Tue, 30 Jul 2024) $

\end{titlepage}

\vfil\break

%%%%%%%%%%%%%%%%%%%%%%%%%%%%%%%%%%%%%%%%%%%%%%%%%%%%%%%%%%%%%%%%

\section{Introduction}    \label{s:intro}

For any supersymmetric theory, the moduli appear as massless fields
transforming in some group representation. This restricts the
holonomy of the moduli space $\cM$. Given enough supersymmetry, this, in
turn, forces the moduli space to be, at least locally, a symmetric
space $\cM=G/K$, where $K$ is a maximally compact subgroup of a semi-simple
Lie group $G$. (Typically, of course, the correct global statement
requires us to further quotient by a discrete group $\Gamma$ to obtain
the true moduli space $\Gamma\backslash G/K$.) 

Since such symmetric spaces are classified, this has proven to be a
very powerful weapon in analyzing moduli spaces and thus these
supersymmetric theories in general. In every dimension, we have a
choice of allowed numbers of supersymmetries and so we naturally go
through many of the possibilities of symmetric spaces. We refer to
\cite{Sez:survey} for a comprehensive review of this.

Going between theories, we often see that some moduli spaces appear
as a subspace in others, $\cM\supset\cM_0$. An
example of this is that the moduli space of supergravity theory can
appear as a subspace of a moduli space of a related lower-dimensional theory.
In terms of compactifications, this can be viewed as decompactifying,
say, a circle. This was called ``Group Disintegrations'' by Julia in
\cite{Jul:dis}. We can also take limits in other parameters. For
example, a full type II string theory moduli space can be viewed as
being parametrized by a conformal field theory, a string coupling, and
a choice of Ramond--Ramond moduli. By taking the zero coupling limit
and ignoring the RR moduli, we embed the moduli space of conformal
field theories within the string moduli space.

This larger space $\cM$ can be
described, as a Riemannian manifold, completely in the language of
$\cM_0$ and the extra parameters in the remaining factors. That is,
the ``disintegration'' does not really lose any information.

The tool we use in this paper to provide the general analysis is the
language of {\em parabolic subgroups}.  We have the familiar
decomposition of a semi-simple Lie algebra into a Cartan subalgebra
and root spaces:
\begin{equation}
  \frak{g} = \frak{h} \oplus\bigoplus_{\alpha\in R}\frak{g}_\alpha,
\end{equation}
where $R$ is the set of roots. The semi-simple condition forces the
condition that if $\alpha\in R$ then $-\alpha\in R$.

We can choose an ordering in the root lattice to divide the roots into
positive and negative roots. A {\em parabolic\/} subalgebra of
$\frak{g}$ is obtained by deleting the root spaces associated to some
negative roots. The resulting subalgebra is thus no longer
semi-simple. Exponentiating the algebra results in a {\em parabolic
  subgroup}.

As we will explain, ignoring any global quotienting discrete group,
this gives a decomposition typically of the form
\begin{equation}
  \cM = N \times \R_+ \times \cM_0.  \label{eq:dec1}
\end{equation}
Here $\R_+$ labels the parameter involved in taking the limit to
obtain $\cM_0$, e.g., the radius of the decompactifying circle for
group disintegrations; while $N$ is a specific space of further
parameters lost. The above product is a topological product but not a
metric one. However, following Borel \cite{Borel:decomp}, we can
explicitly put a metric on the right hand side to make (\ref{eq:dec1})
an isometry. (This is not a product metric or even, in general, a
warped product metric.)

This allows for a full description of $\cM$ in the language of
$\cM_0$. In other words, we can go in the opposite direction of
disintegrations.

We give a couple of applications of this procedure. Let $\cM$ be the
moduli space of superconformal field theories corresponding to a
nonlinear $\sigma$-model on a K3 surface. This is the Grassmanian of
spacelike 4-planes in $\R^{4,20}$. One parameter corresponds to the
volume of the K3. Taking the large limit of this parameter yields the
geometric moduli space of volume one Ricci-flat metrics on a K3
surface, $\cM_0$. We will reconstruct $\cM$ back from $\cM_0$ which
allows an explicit identification of the volume parameter in the
language of Grassmanians. This completes an omission in the original
argument of \cite{AM:K3p}, perpetuated in \cite{me:lK3}.

Our other application is to see how the combinatorics of the symmetric
space decompositions we use can be described by manipulations of
restricted Dynkin diagrams. This allows us to systematically go
through the moduli spaces of higher supergravity theories and describe
their interpretations in terms of various compactifications.

A particularly amusing application of this for maximal supergravities
is the way the parabolic construction interprets passing from geometry
to conformal field theory to superstring theory or M-theory as adding
nodes to a Dynkin diagram. The $D$ and $E$-series of Lie algebras are
thus naturally associated to conformal field theories and supergravity
theories. The ``maximality'' of $E$-type Dynkin diagrams then gives a
nice picture of why one cannot (simply) go ``beyond'' M-theory or
below certain dimensions in compactifications.

Using the parabolic decomposition of the moduli space is also very
useful in going to limits in the moduli space. Indeed, following
geodesic paths in the moduli space always corresponds to such a
description. So we have a fairly general picture of all possible limits.

We apply this limit picture to non-maximal supergravities. This
corresponds to removing nodes from ``restricted'' Dynkin diagrams. In
this analysis, we will follow the disintegration direction by starting
in 3 dimensions and decompactifying. We will see that the geometric
interpretations of these compactifications are always naturally
associated to asymmetric orbifolds along the lines of Dabholkar and
Harvey \cite{Dabholkar:1998kv}.

%%%%%%%%%%%%%%%%%%%%%%%%%%%%%%%%%%%%%%%%%%%%%%%%%%%%%%%%%%%%%%%%

\section{The Hyperbolic Plane} \label{s:hyperbolic}

As a simple warm-up example we consider the Poincar{\'e} half-plane
model of hyperbolic space $\mathsf{H}$ in 2 dimensions. This is viewed
as the complex plane $z\in\C$ with $\Im(z)>0$. $\Sl(2,\R)$ acts on
this half-plane by
\begin{equation}
  \left(\begin{smallmatrix}a&b\\c&d\end{smallmatrix}\right)z
 = \frac{az+b}{cz+d}.  \label{eq:sl2RonH}
\end{equation}
We can fix a ``fiducial'' point, $z=i\in\C$, and then the orbit of $\Sl(2,\R)$
acting on this point covers all of $\mathsf{H}$. The stabilizer
of $i$ is easily seen to be the subgroup with elements of the form
\begin{equation}
\begin{pmatrix}\cos(\theta)&-\sin(\theta)\\\sin(\theta)&\cos(\theta)
\end{pmatrix},
\end{equation}
which forms $\SO(2)=\GU(1)$. Note that $\GU(1)$ is the maximal
subgroup of $\Sl(2,\R)$ that is compact. This yields an equivalence
\begin{equation}
  \mathsf{H} = \frac{\Sl(2,\R)}{\GU(1)}.  \label{eq:isom1}
\end{equation}
This equivalence is a diffeomorphism, but we can go further and make
it an {\em isometry\/} if we include metrics.

Consider the following metric on $\mathsf{H}$ in the coordinates of
the upper-half plane:
\begin{equation}
\begin{split}
ds^2 &= \frac{dz\,d\bar z}{\Im(z)^2}\\
&= \frac{dx^2+dy^2}{y^2}.  \label{eq:Hyperbolic}
\end{split}
\end{equation}
This has constant negative Gaussian curvature. Furthermore, it is
straight-forward to see that it is invariant under the $\Sl(2,\R)$
action (\ref{eq:sl2RonH}).

Now let's construct a natural metric more directly in the language of
the coset $\Sl(2,\R)/\GU(1)$. Our fiducial point $i$ is taken to
$\sigma=x+iy\in\mathsf{H}$ by the upper-triangular
$\Sl(2,\R)$ matrix
\begin{equation}
  \begin{pmatrix}y^{\frac12}&xy^{-\frac12}\\0&y^{-\frac12}\end{pmatrix}.
\end{equation}
Upper triangular matrices of determinant one with positive diagonal
entries form a parabolic subgroup $P$ of $\Sl(2,\R)$. Indeed, up to
equivalence by conjugation, this is the unique connected nontrivial
parabolic subgroup.

Note that any element of $\Sl(2,\R)$ can be written uniquely (if we assume
$y^{\frac12}>0$) as
\begin{equation}
  \begin{pmatrix}y^{\frac12}&xy^{-\frac12}\\0&y^{-\frac12}\end{pmatrix}
\begin{pmatrix}\cos(\theta)&-\sin(\theta)\\\sin(\theta)&\cos(\theta)
\end{pmatrix}
\end{equation}
Thus $P$ is in bijection with $\mathsf{H}$. Elements of $P$ can be
factorized as
\begin{equation}
\begin{pmatrix}1&x\\0&1\end{pmatrix}
\begin{pmatrix}y^{\frac12}&0\\0&y^{-\frac12}\end{pmatrix}.
\end{equation}
Let $N$ be the group of elements of the form of the first factor and
$A$ the group corresponding to the second factor.
Since $x\in\R$ and $y\in\R_+$, this parabolic
subgroup parametrization yields set bijection
\begin{equation}
\mathsf{H} = \R \times \R_+.  \label{eq:HRR}
\end{equation}

The corresponding Lie algebras $\mathfrak{n}$ and $\mathfrak{a}$ are given,
respectively, by
\begin{equation}
\begin{split}
  \begin{pmatrix}1&x\\0&1\end{pmatrix}
 &= \exp\begin{pmatrix}0&x\\0&0\\\end{pmatrix}\\
  \begin{pmatrix}y^{\frac12}&0\\0&y^{-\frac12}\end{pmatrix}
 &= \exp\begin{pmatrix}a&0\\0&-a\\\end{pmatrix},\\
\end{split}
\end{equation}
where $y=\exp(2a)$.

Consider the adjoint action of $A$ on $N$:
\begin{equation}
\begin{pmatrix}y^{-\frac12}&0\\0&y^{\frac12}\end{pmatrix}
\begin{pmatrix}1&x\\0&1\end{pmatrix}
\begin{pmatrix}y^{\frac12}&0\\0&y^{-\frac12}\end{pmatrix}
=
\begin{pmatrix}1&y^{-1}x\\0&1\end{pmatrix}.
\end{equation}
Since $x$ also parametrizes the Lie algebra $\mathfrak{n}$, we have an induced adjoint
action
\begin{equation}
\Int(a)x = y^{-1}x,
\end{equation}
and thus on one-forms in the Lie algebra
\begin{equation}
\Int(a)^*dx = y^{-1}dx.
\end{equation}

We can then reconstruct the hyperbolic metric above by noting that
\begin{equation}
  4(da)^2 + \left(\Int(a)^*dx\right)^2 = \frac{dx^2+dy^2}{y^2}.
  \label{eq:hpm}
\end{equation}
The $\Sl(2,\R)$-invariant metric is defined only up to an
overall scale, but the relative scaling between the two terms of the
left of (\ref{eq:hpm}) is significant. We will carefully analyze this
in section \ref{ss:metric}.

Anyway, this shows that (\ref{eq:HRR}) can be made an isometry but that the
resulting metric does not fully respect the product structure in the
sense that we need to add the $\Int(a)$ action.

%%%%%%%%%%%%%%%%%%%%%%%%%%%%%%%%%%%%%%%%%%%%%%%%%%%%%%%%%%%%%%%%%%%

\section{Parabolic Subgroups.}   \label{s:para}

We now generalize the previous section using some standard methods
\cite{Knapp:beyond}.  Let $\frak{g}$ be a semi-simple real Lie group,
with its familiar root space decomposition
\begin{equation}
  \frak{g} = \frak{h}
  \oplus\bigoplus_{\alpha}\frak{g}_\alpha,  \label{eq:rootdecomp}
\end{equation}
where the sum is over all the roots. The Cartan subalgebra
$\frak{h}$ is a maximal abelian subalgebra, and each root space
$\frak{g}_\alpha$ is one-dimensional. Choosing an order, we may divide
the set of roots into positive roots and negative roots in the usual way.

Any real form of a semi-simple Lie algebra admits a ``Cartan
involution'' $\theta$, which may always be chosen to be
$\theta(X)=-X^\dagger$. The important property is that, if $B$ is the
Killing form, we can define a new inner product:
\begin{equation}
B_0(X,Y) = -B(X,\theta(Y)),
\end{equation}
which is a symmetric {\em positive definite\/} form on $\frak{g}$.

We have a ``Cartan decomposition''
\begin{equation}
\frak{g} = \frak{k}\oplus \frak{q},
\end{equation}
into $\pm1$ eigenspaces of $\theta$ respectively. We think of
$\frak{k}$ as the ``compact'' part of $\frak{g}$ as it exponentiates
to a compact subgroup $K\subset G$. Indeed, this is a maximal compact subgroup.

We would like to redo the root decomposition (\ref{eq:rootdecomp})
again but with respect to a noncompact version of a Cartan subalgebra.
So let $\frak{a}$ be a maximal abelian subalgebra of $\frak{q}$.  We
can then decompose $\frak{g}$ into eigenspaces of $\frak{a}$:
\begin{equation}
\frak{g} = \frak{g}_0 \oplus
\bigoplus_{\lambda\in\Sigma}\frak{g}_\lambda,
\end{equation}
where $\Sigma$ are the nonzero {\em restricted roots\/} living in
$\frak{a}^*$. If $\frak{g}$ is the {\em split\/} (i.e., maximally
noncompact) form then, and only then, we can set $\frak{a}=\frak{h}$
and the roots and restricted roots coincide. At the other extreme, the
{\em compact\/} form of $\frak{g}$ has no restricted roots.

The space $\frak{g}_0$ is the eigenspace of $\frak{a}$ with zero
eigenvalue and obviously includes $\frak{a}$ itself. We define the
subalgebra $\frak{m}$ by writing
\begin{equation}
 \frak{g}_0 = \frak{a}\oplus\frak{m}.
\end{equation}

Fix an ordering to define {positive\/} restricted roots
$\Sigma^+$. Amongst those we can define {\em simple\/} roots and
construct a Dynkin diagram in the usual way. We'll call this the {\em
restricted\/} Dynkin diagram. The rules for restricted Dynkin diagrams
are very similar to, but not quite the same as usual Dynkin
diagrams. In particular, one has the familiar rules

\begin{enumerate}
  \item If $\alpha$ is a root of a semi-simple Lie algebra, then $2\alpha$ is
    not a root.
  \item $\dim\frak{g}_{\alpha}=1$ for any root $\alpha$.
\end{enumerate}

This is no longer the case for restricted roots. The classification of
restricted Dynkin diagrams for simple Lie algebras is therefore
slightly relaxed. In addition to the usual cases, there is one more
family which is like a union of $B_n$ and $C_n$ and is denoted
$(BC)_n$. This consists of roots $\pm L_i\pm L_j$, $\pm L_i$ and $\pm
2L_i$. For this we'll use the Dynkin diagram:
\begin{equation}
(BC)_n:\qquad \dynkin[edge length=8mm,arrows=false]{C}{o.ooO}
\end{equation}

Since $\dim\frak{g}_{\alpha}$ need not be one, we can have more
interesting root spaces. We will label the nodes in the Dynkin diagram
accordingly as we explain shortly.

Next define the subalgebra
\begin{equation}
\frak{n} = \sum_{\lambda\in\Sigma^+}\frak{g}_\lambda.
\end{equation}
One can then prove we have a vector space decomposition
\begin{equation}
\frak{g} = \frak{n}\oplus\frak{a}\oplus\frak{k}.
\end{equation}
This can be exponentiated to form the {\em Iwasawa decomposition\/} of
the group
\begin{equation}
  G = N\times A\times K.
\end{equation}
This product form is not an isomorphism of groups. What is true is that the set map
$N\times A\times K\to G$ given by $(n,a,k)\to nak$ is a
diffeomorphism and onto.

The subalgebra
\begin{equation}
\frak{p}_m = \frak{n}\oplus\frak{a}\oplus\frak{m},
\end{equation}
is called the {\em minimal parabolic subalgebra\/} of $\frak{g}$.

We want to find bigger parabolic subalgebras lying between
$\frak{p}_m$ and $\frak{g}$ itself. To this end, let us relabel things
a bit. Denote $\frak{n}$, $\frak{a}$ and $\frak{m}$ above by $\frak{n}_m$,
$\frak{a}_m$ and $\frak{m}_m$ respectively. We will then find new
$\frak{n}$, $\frak{a}$, $\frak{m}$'s for our new non-minimal parabolic
subalgebras.

Let $\Pi$ be the set of {\em simple\/} restricted roots. Choose
$\Pi'\subset\Pi$. Then define
\begin{equation}
  \Gamma = \Sigma^+ \cup \{\alpha\in\Sigma: \alpha\in\Span\Pi'\}.
\end{equation}
So $\Gamma$ is the set of all positive roots $\Sigma^+$ with some
negative roots added. Let $\Gamma^-$ denote these new negative
roots. Now define a smaller $\frak{a}$:
\begin{equation}
\frak a = \bigcap_{\alpha\in\Gamma^-}\ker\alpha\subset\frak{a}_m, \label{eq:akerB}
\end{equation}
and a smaller $\frak{n}$:
\begin{equation}
\frak n =
\bigoplus_{\alpha\in\Sigma^+,-\alpha\not\in\Gamma^-}\frak{g}_\alpha,
    \label{eq:nsum}
\end{equation}
and a bigger $\frak{m}$:
\begin{equation}
\frak m = \frak a^\perp\oplus \frak{m}_m \oplus
   \bigoplus_{\alpha\in\Gamma\cap-\Gamma}\frak{g}_\alpha,
\end{equation}
where $\frak{a}^\perp$ is the orthogonal complement of $\frak{a}$ in
$\frak{a}_m$.

The algebras $\frak{n}$, $\frak{a}$ and $\frak{m}$ then generate a
parabolic subalgebra $\frak{p}$ of $\frak{g}$. All parabolic
subalgebras lying between $\frak{p}_m$ and $\frak{g}$ can be written
in this form. We have a vector space decomposition
\begin{equation}
 \frak{p} = \frak{n}\oplus\frak{a}\oplus\frak{m},
\end{equation}
called the {\em Langlands decomposition\/} of $\frak{p}$.

If $\Pi'$ is given by a single simple root $\alpha$, let
$\frak{m}_\alpha$ denote the corresponding subalgebra $\frak{m}$. We
will label each node in the Dynkin diagram by this corresponding
$\frak{m}_\alpha$ following \cite{Knapp:beyond}. Since $\sl(2,\R)$ and
$\sl(2,\H)$ occur frequently, we will label these nodes simply by $\R$
and $\H$ respectively. Note that for the split form, each node is
labeled by $\R$.

Let $N$, $A$ and $M$ be the corresponding subgroups of $G$ given by
exponentiation. Let $P=NAM$ be the subset of elements of the form
$nam$. Then one can show $P$ is a subgroup of $G$ and is thus a
parabolic subgroup.

$M$ and $A$ commute, and $MA$ normalizes $N$ in $P$. So $MA$ has an
adjoint action, $\Int$, on the Lie algebra $\frak{n}$. In particular,
$\frak{n}$ is a representation of $\frak{m}$.

One can also show $G=PK$. That is, any element of $G$ can be written
as $pk$, $p\in P$, $k\in K$. This writing might not be unique
however --- one can have nontrivial $K_P=P\cap K$. However, if we write
$g=namk$, then $n$ and $a$ are uniquely determined by $g$.

So, when we compute the quotient $G/K$, this factorizes set-wise as
\begin{equation}
G/K = N\times A \times Z,  \label{eq:NAZ}
\end{equation}
where $Z=M/K_P$. From now on, to make the equations a little more
readable and to fit in with the rest of the physics literature, we
will often write right quotients vertically $\frac GK$ in this factorization.

\subsection{The Metric on $G/K$}  \label{ss:metric}

The inner product $B_0$ on $\frak{g}$ can be translated over $G$
by the (left) group action to produce the bi-invariant metric on $G$. The
decomposition $\frak{g}=\frak{k}\oplus\frak{q}$ is orthogonal with
respect to $B_0$ and thus we naturally have a metric on $G/K$
given by the metric $B_0$ restricted to $\frak{q}$.

When we do our parabolic decomposition this metric will split into
orthogonal parts related to $\frak{n}$, $\frak{a}$ and $\frak{m}$. Let
us analyze this following \cite{Borel:decomp}.\footnote{The reference
  \cite{Borel:decomp} considers right-invariant metrics, left
  quotients $K\backslash G$ and group factorizations in the reverse
  directions to this paper.}

Let us assume $\frak{a}$ is one-dimensional (as it usually will be for
our examples) and is given by $tR$, for some generator $R\in\frak{h}$,
and $t\in\R$. Next, let $\lambda=\exp(t)$ parametrize the
corresponding group $A$. The metric for the $A$ part is then
\begin{equation}
  da^2 = B(R,R)dt^2 = B(R,R)\frac{d\lambda^2}{\lambda^2}.
\end{equation}

The $\frak{n}$ algebra is nilpotent and the Killing form of
$\frak{g}$ is zero when restricted to it. However, we have
positive-definite metric on $\frak{n}$ when we consider $B_0$. Let
$dn^2$ denote this metric.  An element $n\in\frak{n}$ decomposes
\begin{equation}
  n = \ff12(n+\theta(n)) +
  \ff12(n-\theta(n))\in\frak{k}\oplus\frak{q}.
\end{equation}
Thus the inner product for $\frak{n}$ when we project into $\frak{q}$
is
\begin{equation}
  \begin{split}
    (n_1,n_2)_{\frak{q}} &=
                B_0\left(\ff12(n_1-\theta(n_1)),\ff12(n_2-\theta(n_2))\right)\\
    &= \ff12 B_0(n_1,n_2).
  \end{split}
\end{equation}

Let $dz^2$ denote the invariant metric on the space $Z=M/(K\cap
P)$. The action $\Int(am)$ for $a\in A$ and $m\in M$ on $N$ induces an action
$\Int^*$ on the metric. We have therefore shown \cite{Borel:decomp}:

\begin{prop}  \label{prop:ds2}
For our parabolic decomposition of the quotient $G/K$ in
(\ref{eq:NAZ}) the invariant metric can be written
\begin{equation}
  ds^2 = \ff12\Int(am)^*dn^2 + da^2 + dz^2. \label{eq:Borelmetric}
\end{equation}
\end{prop}

We can be quite explicit about the $\Int(am)^*$ action. If we consider
decomposing the adjoint representation of $\frak{g}$ under the action
of the subalgebra $\frak{m}\subset\frak{g}$ we explicitly see how
$\frak{n}$ forms a representation of $\frak{m}$. This yields the
$\Int(m)^*$ action.

Let $\alpha$ be a positive root used to build $\frak{n}$
and let $X_\alpha\in\frak{g}_\alpha$, $Y_\alpha=-\theta(X_\alpha)\in\frak{g}_{-\alpha}$
and $H_\alpha=[X_\alpha,Y_\alpha]\neq0$. We have
\begin{equation}
  [H_\alpha,X_\alpha]=\alpha(H_\alpha)X_\alpha.
\end{equation}
It is common to normalize the choice of $X_\alpha$ so that
$\alpha(H_\alpha)=2$ but we will not always do this. We claim
\begin{prop}  \label{prop:BXaXa}
\begin{equation}
  B_0(X_\alpha,X_\alpha) = \frac{\alpha(H_\alpha)}{(\alpha,\alpha)}.
\end{equation}
\end{prop}
To prove this, note that the Killing form gives a natural isomorphism
$\frak{a}_m^*\cong \frak{a}_m$. Under this, a restricted root $\alpha$ is mapped to
$T_\alpha$ defined by
\begin{equation}
  B(T_\alpha,H) = \alpha(H), \quad\hbox{for all $H\in\frak{a}_m$}.
\end{equation}
But
\begin{equation}
  \begin{split}
    B(H_\alpha,H) &= B([X_\alpha,Y_\alpha],H)\\
                  &= B(X_\alpha,[Y_\alpha,H])\\
                  &= \alpha(H)B(X_\alpha,Y_\alpha).
  \end{split}
\end{equation}
So
\begin{equation}
  B_0(H_\alpha,H_\alpha) = \alpha(H_\alpha) B_0(X_\alpha,X_\alpha).
\end{equation}
It follows that
\begin{equation}
  T_\alpha = \frac{\alpha(H_\alpha) H_\alpha}{B(H_\alpha,H_\alpha)}.
\end{equation}
We can also use this isomorphism to copy the Killing form over to an
inner product on the root space. Then
\begin{equation}
  \begin{split}
    (\alpha,\beta) &= \beta(T_\alpha)\\
                   &=
                     \frac{\alpha(H_\alpha)\beta(H_\alpha)}{B(H_\alpha,H_\alpha)}\\
  \end{split}
\end{equation}
The proposition follows.

In the applications we consider we will generally have
$B_0(X_\alpha,X_\alpha) = 1$.

If $\frak{a}$ is one-dimensional we can again use the $R\in\frak{h}$
generator. Then $R$ acts on $X_\alpha$ with eigenvalue
$\alpha(R)$. This is the action of $\Int(a)$.  Assuming
$B_0(X_\alpha,X_\alpha) = 1$, if we parametrize elements of
$\frak{g}_\alpha\subset\frak{n}$ as $x_\alpha X_\alpha$, then this
leads to a metric,
\begin{equation}
  ds^2 = \sum_\alpha\frac{1}{2\lambda^{2\alpha(R)}}\Int(m)^*dx_\alpha^2 +
  B_0(R,R)\frac{d\lambda^2}{\lambda^2}+dz^2,
\end{equation}
where the $\alpha$ sum is that of (\ref{eq:nsum}).
In particular, if we restrict to $m=1$, i.e., the image of the
identity in $Z$, we have
\begin{equation}
  ds^2|_{m=1} = \sum_\alpha\frac{dx_\alpha^2}{2\lambda^{2\alpha(R)}} +
  B_0(R,R)\frac{d\lambda^2}{\lambda^2}.   \label{eq:metricm=1}
\end{equation}
Note that rescaling $R$ is equivalent to replacing $\lambda$ by a
power of itself.

A key idea is that we are thus able to construct the metric
(\ref{eq:Borelmetric}) purely from a knowledge of which representation
$\frak{n}$ forms for $\frak{m}$, and how $\frak{a}$ acts, i.e., the
analogue of the $y^{-2}$ factor in (\ref{eq:hpm}).

Let us check this is consistent with the hyperbolic plane. Here,
$\frak{g}\cong\sl(2,\R)$ generated by the usual matrices
\begin{equation}
  H=\begin{pmatrix}1&0\\0&-1\end{pmatrix},\quad
  X=\begin{pmatrix}0&1\\0&0\end{pmatrix},\quad
  Y=\begin{pmatrix}0&0\\1&0\end{pmatrix}.
\end{equation}
The subalgebra $\frak{k}$ is generated by $X-Y$ while $\frak{q}$ is
generated by $H$ and $X+Y$. So $\frak{m}$ is trivial, $\frak{a}$ is
generated by $R=H$ and $\frak{n}$ is generated by a single positive
root $\alpha$ associated to $X$.

Now $H_\alpha=H$ and $B_0(R,R)=\alpha(R)=(\alpha,\alpha)=2$ so we have
\begin{equation}
  ds^2 = \frac{dx^2}{2\lambda^4} + \frac{2d\lambda^2}{\lambda^2}.
\end{equation}
Putting $y=\lambda^2$, we recover (\ref{eq:Hyperbolic}) up to an
(irrelevant) overall factor of 2.

\subsection{Geodesic Limits}  \label{ss:limits}

Finally, in this general discussion let us consider the idea of
finding a limit by following a path in the moduli space to infinite
distance. Clearly we could follow some crazy space-filling curve and
get a quite arbitrary result. So let us limit attention to following {\em
  geodesics}. We have the following well-known \cite{Helgason:Lie}
\begin{prop}
  Any complete geodesic on a symmetric space $G/K$ is given as an
  orbit of a one-parameter subgroup of the isometry group $G$.  
\end{prop}
The tangent space at the ``origin'', i.e., the image of $1\in G$ is
given by $\frak{q}$. So if we start at this origin, we need to pick
some element $Q\in\frak{q}$ and follow its exponential $\exp(tQ)$ to
go along the geodesic. Since our space is homogeneous, we can
translate this statement to any starting point.

Any element of $\frak{g}$ can be conjugated into a Cartan subalgebra
$\frak{h}$. So, fixing the Cartan decomposition, $Q$ may be conjugated
into $\frak{a}_m$.

So now we have a parabolic decomposition by defining
$\frak{a}\subset\frak{a}_m$ to be generated by $Q$. Since $\frak{a}$
is one-dimensional we must have that $\Pi'$ is given by removing one
element from $\Pi$.

This shows
\begin{prop}
  Any geodesic generated by a Lie algebra element $Q\in\frak{q}$
  yields a parabolic decomposition (\ref{eq:NAZ}) where $A$ is
  one-dimensional and given by the completion of this geodesic. This
  corresponds to deleting one node from the restricted Dynkin diagram.
\end{prop}

So, actually, considering geodesics, our parabolic decomposition
method gives the general picture for limiting paths going off
to infinity, where $\lambda\in A\cong\R_+$ is playing the role of the
limiting parameter. This means finding limits along geodesics
always corresponds to removing nodes from the Dynkin diagram.

We emphasize that the logic here is that you choose a limit and that
yields a parametrization. Suppose, instead, you already have a
parametrization of the moduli space in terms of a parabolic
decomposition. That is, you have already decided how to label each
point in the moduli space by size, $B$-field, etc. Now when you pick
your element $Q$ to generate a geodesic, it does not need to be an
element of $\frak{a}$. So now your limit is a combination of changing
the size while varying the $B$-field, for example.

That said, you can always conjugate everything by some element of $G$
to move $Q$ into $\frak{a}_m$. So this arbitrary limit can be
understood as a conjugation of a parabolic decomposition.
  
\section{The K3 Moduli Space}

For our first example we will consider the moduli space of $N=(4,4)$ conformal field
theories with $c=6$. This will result in the isometry\footnote{Since
  our analysis in this paper is always local we will not be careful
  distinguishing between $\SO(n)$ and $\GO(n)$, etc.} 
\begin{equation}
    \frac{\SO(4,20)}{\SO(4)\times\SO(20)} =
    \R^{22}\times\R_+ \times\frac{\SO(3,19)}{\SO(3)\times\SO(19)}. \label{eq:K3a}
\end{equation}
As explained in \cite{AM:K3p,me:lK3} this gives, up to a global
discrete quotient, the geometric
interpretation of the conformal field theories in terms of a $B$-field
and a K3 metric. Let us analyze this carefully using the parabolic
decomposition language. This will allow us to justify an assertion
about the identification of parameters made in \cite{me:lK3}.

\subsection{The metric}  \label{ss:K31}

In order to interpret (\ref{eq:K3a}) as an isometry we need to compute
the metric (\ref{eq:Borelmetric}). Consider the symmetric space
\begin{equation}
  \frac{\SO(p,q)}{\SO(p)\times\SO(q)}  \label{eq:sopq}
\end{equation}
where we will assume $p\leq q$. 

Let $\frak{g}=\so(p,q)$. The signature $(p,q)$ bilinear form we
preserve can be chosen to be the one associated to the block form
\begin{equation}
\Xi = \begin{pmatrix} 0&I&0\\I&0&0\\0&0&-I\end{pmatrix},
\end{equation}
where we use blocks of size $p,p,q-p$. An element $X\in\frak{g}$ satisfies
$X^T\Xi+\Xi X=0$ and has general block form
\begin{equation}
  X = \begin{pmatrix}\mathsf{A}&\mathsf{B}&\mathsf{D}\\\mathsf{C}&
      -\mathsf{A}^T&\mathsf{E}\\\mathsf{E}^T&\mathsf{D}^T&\mathsf{G}
      \end{pmatrix},
\end{equation}
where $\mathsf{B}$, $\mathsf{C}$ and $\mathsf{G}$ are skew-symmetric.

Let $\theta(X)=-X^T$ be the Cartan involution. We have a decomposition
\begin{equation}
\frak{g} = \frak{k}\oplus \frak{q},
\end{equation}
into $\pm1$ eigenspaces of $\theta$. The subalgebra $\frak{k}$ is
given by $\mathsf{A}$ and $\mathsf{G}$, both skew-symmetric,
$\mathsf{B}=\mathsf{C}$ and $\mathsf{D}=-\mathsf{E}$. This forms the
maximal compact subalgebra $\so(p)\oplus\so(q)$.

The abelian subalgebra $\frak{a}_m$ has a basis $H_i$, $i=1,\ldots,p$,
given by a single nonzero entry for $\mathsf{A}$ in the $i$th position
on the diagonal:
\begin{equation}
  \mathsf{A} = \left(\begin{smallmatrix}0\\&\ddots\\&&0\\&&&1\\&&&&0\\&&&&&\ddots
               \end{smallmatrix}\right).
\end{equation}
We define a basis for the restricted root space, $L_i$, by declaring
$L_i(H_j)=\delta_{ij}$. We then have the familiar roots
$\pm L_i\pm L_j$, $i\neq j$ for this algebra each of multiplicity
one. Assuming $q>p$ we also have roots $\pm L_i$, each of multiplicity
$q-p$. An ordering can then be chosen to yield the restricted simple
roots $L_1-L_2,L_2-L_3,\ldots,L_{p-1}-L_p,L_p$ and a restricted Dynkin
diagram of type $B_p$:
\begin{equation}
  \dynkin[edge
  length=8mm, label directions={,,,right},
  labels={\R,\R,\R,{\so(1,q-p+1)}}]{B}{oo.oo}
\end{equation}

Let $w$ be a null vector in $\R^{p,q}$. In the basis above let it be
$w=\left(\begin{smallmatrix}1\\0\\\vdots\end{smallmatrix}\right)$. Consider
the subgroup $P\subset\SO(p,q)$ that preserves the ray spanned by $w$. The
first column of the matrix of such an element must have all zeroes below
the first row. So the same is true for the Lie algebra of this
subgroup. We thus restrict the algebra element $X$ above by saying
\begin{itemize}
\item The first column of $\mathsf{A}$ below the first row is zero.
\item The first column of $\mathsf{C}$ is zero.
\item The first row of $\mathsf{E}$ is zero.
\end{itemize}
In terms of roots, we are therefore deleting the root spaces of
$-L_1+L_i$, $-L_1-L_i$ and $-L_1$ respectively for $i=2,\ldots,p$.

The Lie algebra of $P$ is a parabolic subalgebra $\frak{p}\subset\so(p,q)$. In the
language of the previous sections this is given by
\begin{equation}
\Pi' = L_2-L_3,L_3-L_4,\ldots,L_{p-1}-L_p,L_p.
\end{equation}
This corresponds to obtaining $\Pi'$ from $\Pi$ by deleting the
leftmost node:
\begin{equation}
  \dynkin[edge
  length=8mm, label directions={,,,right},
  labels={\R,\R,\R,{\so(1,q-p+1)}}]{B}{xo.oo}
\end{equation}
So $\frak{a}$ is one-dimensional corresponding to $R=H_1$.
The subgroup $A$ has elements
\begin{equation}
e^a = \left(\begin{array}{cccc|cccc|ccc}
\lambda&&&&&&&&\\
&1&&&&&&&\\
&&\ddots&&&&&&\\
\hline
&&&&\lambda^{-1}&&&&\\
&&&&&1&&&\\
&&&&&&\ddots&&\\
\hline
&&&&&&&&1\\
\end{array}
\right).  \label{eq:K3alambda}
\end{equation}

The nilpotent subalgebra $\frak{n}$ is given by the positive
restricted roots $L_1\pm L_i$ and $L_1$. An element is thus given by
(we choose signs and subscripts judiciously!)
\begin{equation}
n = \left(\begin{array}{cccc|cccc|cccc}
0&-B_p&-B_{p+1}&\cdots & 0&-B_1&-B_2&\cdots & B_{2p-1}&B_{2p}&\cdots&B_{p+q-2}\\
0&0&0&&B_1&0&0&&0&0&&0\\
0&&&&B_2&&&&&\\
\vdots&&&&\vdots&&&&\vdots\\
0&&&&B_{p-1}&&&&&\\
\hline
0&0&0&\cdots&0&&&&0\\
0&&&&B_p&&&&0\\
0&&&&B_{p+1}&&&&0\\
\vdots&&&&\vdots&&&&\vdots\\
\hline
0&&&&B_{2p-1}&&&&0\\
0&&&&B_{2p}&&&&0\\
\vdots&&&&\vdots&&&&\vdots\\
\end{array}
\right).  \label{eq:nbK3}
\end{equation}

The subalgebra $\frak{m}$ is given by $\so(p-1,q-1)$. The action of
$\Int(m)$ on $\frak{n}$ is simply to rotate the $(p+q-2)$-dimensional
vector $B$ given by the above matrix.

Now we determine the metric on the decomposition (\ref{eq:K3a}). Let
$X_i$, $i=1,\ldots,p+q-2$, denote the matrix in (\ref{eq:nbK3}) with
$B_j=\delta_{ij}$. So the $X_i$'s are a basis for $\frak{n}$.
For $i=1,\ldots,2p-2$ we have
\begin{equation}
  H_{\alpha_i}=[X_i,X_i^T] = H_1\pm H_k,\quad \alpha_i(H_{\alpha_i}) = 2,
\end{equation}
for some $k\neq1$. On the other hand, for $i=2p-1,\ldots,p+q-2$ we have 
\begin{equation}
  H_{\alpha_i}=[X_i,X_i^T] = H_1,\quad \alpha_i(H_{\alpha_i}) = 1.
\end{equation}
In either case we have $B_0(X_\alpha,X_\alpha)=1$ from proposition
\ref{prop:BXaXa}.

So proposition \ref{prop:ds2} yields the metric, for the $m=1$ slice,
\begin{equation}
  ds^2|_{m=1} = \frac{dB^2}{2\lambda^2} +
  \frac{d\lambda^2}{\lambda^2},  \label{eq:K3m1}
\end{equation}
where $dB^2=\sum_idB_i^2$. Thus we are expanding the $B$-field in terms of
a basis of 2-forms $B_i$ which are orthogonal with respect to our metric.

Now this is supposed to agree with the Zamolodchikov metric of
conformal field theories on a K3 surface. This was given in
\cite{CHS:metric} but we don't really need to know the details. The
key point is that the non-linear sigma model can be written purely in
terms of the combination $g_{\mu\nu}+B_{\mu\nu}$. Because of this, the
Zamolodchikov metric for a change $\delta g_{\mu\nu}$ and $\delta B_{\mu\nu}$
is proportional to
\begin{equation}
  \delta g_{\mu\nu}\delta g_{\xi\zeta} + \delta B_{\mu\nu}\delta B_{\xi\zeta}.  \label{eq:Zam}
\end{equation}
As is well known, the Ricci-flat metric data can be recast in terms of
a complex structure and a K\"ahler form.  The K\"ahler form $J$ and
$B$-field $B$ are both 2-forms and we can think of the above, for
fixed complex structure, in the orthonormal cohomology basis as
$dJ^2+dB^2=\sum_i(dJ_i^2+dB_i^2)$.

Let us write $J=j\hat J$ where $j$ is scalar so that
$\int J^2=j^2$. Then $\hat J$ is some normalized version of the K\"ahler
form. But then the $Z$ factor of our moduli space,
$\SO(3,19)/(\SO(3)\times\SO(19))$, is known to be the moduli space of
complex structures and K\"ahler forms for a fixed volume
\cite{Tod:inv,Mor:Katata,Kobayashi-Todorov}.

Rescaling the metric to produce a change in $j$ is orthogonal to any
of the moduli preserving the volume. This can be seen as follows. The
metric on the space of K\"ahler forms is simply induced by the natural
inner product on the space of $(1,1)$-forms \cite{Cand:mir}:
\begin{equation}
  a\wedge *b  = (a,b)\omega, 
\end{equation}
where $\omega$ is the volume form. The K\"ahler form itself is
self-dual $*J=J$. So, if we deform the K\"ahler form by $\delta J$ so as
to keep the volume constant, we have $(J,\delta J)=0$. Now, volume
deformations can be seen as rescaling $J$, so such a deformation is
orthogonal to the volume-preserving $\delta J$.

If we look at a component $J_i=j\hat J_i$ of the cohomology class, we
have a deformation
\begin{equation}
  dJ_i = \hat J_i\,dj + j\,d\hat J_i.
\end{equation}
Squaring and summing over $i$ we therefore have a K\"ahler form
contribution to the metric
\begin{equation}
  \frac{dJ^2}{j^2} = \frac{dj^2}{j^2} + d\hat J^2. \label{eq:K3mj}
\end{equation}
Now $d\hat J^2$ is part of $dz^2$ in the parabolic decomposition. We
therefore get agreement between (\ref{eq:K3m1}) and (\ref{eq:K3mj}) by
setting
\begin{equation}
  \lambda^2=\ff12j^2=\ff12\int J^2.
\end{equation}
Taken altogether, we have proven the parametrization as follows:
\begin{prop}
  The decomposition (\ref{eq:K3a}) is an isometry between the
  moduli space of conformal field theories on K3 surface with the
  Zamolodchikov metric and the triple $(Z,A,N)$, where $Z$ specifies a
  Ricci flat metric on a K3 surface of volume 1, $A$ specifies the
  volume and $N$ specifies the $B$-field where the metric is given by
  \begin{equation}
    ds^2 = dz^2 + \frac{\Int(m)^*(dB^2)+dj^2}{j^2},
  \end{equation}
  and the volume is $\ff12j^2=\lambda^2$. $N$ is given by $\exp(n)$ in
  terms of $B$ as given in (\ref{eq:nbK3}) and $A=\exp(a)$ in terms of
  $\lambda$ as given in (\ref{eq:K3alambda}). The action of $\Int(m)$
  is to rotate the vector $B$.
\end{prop}

\subsection{The Grassmanian description}

The symmetric space (\ref{eq:sopq}) can be viewed as the Grassmanian
of spacelike $p$-planes in $\R^{p,q}$. Let us view the parametrization
of the moduli space in this language.

First we need an explicit form of $N=\exp(n)$. Note that $n^2$ has a single
entry, $-B^2$, using our inner product $\Xi$,
\begin{equation}
  B^2 = B_1B_{p}+B_2B_{p+1}+\ldots+B_{p-1}B_{2p-2}-B_{2p-1}^2-\ldots,
\end{equation}
in the $(1,p)$ slot, and $n^3=0$. So, the group element is
\begin{equation}
e^n = \left(\begin{array}{cccc|cccc|ccc}
1&-B_p&-B_{p+1}&\cdots & -\ff12B^2&-B_1&-B_2&\cdots & B_{2p-1}&B_{2p}&\cdots\\
0&1&0&&B_{1}&0&0&&0&0\\
0&0&1&&B_{2}&&&&&\\
\vdots&&&&\vdots&&&&\vdots\\
\hline
0&0&0&\cdots&1&0&0&0&0\\
0&&&&B_{p}&1&0&&0\\
0&&&&B_{p+1}&0&1&0&0\\
\vdots&&&&\vdots&&&&\vdots\\
\hline
0&&&&B_{2p-1}&&&&1\\
0&&&&B_{2p}&&&&0\\
\vdots&&&&\vdots&&&&\vdots\\
\end{array}
\right).
\end{equation}

Let us explicitly produce a spacelike $p$-plane for the Grassmanian interpretation.
We can choose a ``fiducial'' fixed spacelike $p$-plane as the column space of
the matrix
\begin{equation}
\Pi_0 = \begin{pmatrix}I\\I\\0\end{pmatrix}.
\end{equation}
So we rotate this by an element in $NA$ to give
\begin{equation}
e^ne^a\Pi_0 = \left(\begin{array}{cccc}
\lambda-\ff12\lambda^{-1}B^2&-B_1-B_{p}&-B_2-B_{p+1}&\cdots\\
\lambda^{-1}B_1&1&0&\cdots\\
\lambda^{-1}B_2&0&1&\cdots\\
\vdots&&&\ddots\\
\hline
\lambda^{-1}&0&0&\cdots\\
\lambda^{-1}B_p&1&0&\cdots\\
\lambda^{-1}B_{p+1}&0&1&\cdots\\
\vdots&&&\ddots\\
\hline
\lambda^{-1}B_{2p-1}&0&0&\cdots\\
\lambda^{-1}B_{2p}&0&0&\cdots\\
\vdots&&&\ddots\\
\end{array}
\right),
\end{equation}
whose column space is now our desired spacelike $p$-plane, $\Pi$.

To parametrize this Grassmanian, the recipe in \cite{me:lK3} is to now
find $\Sigma'=\Pi\cap w^\perp$. Clearly only the first column above has
nonzero inner product with $w$ so $\Sigma'$ is the span of the
remaining columns. We now define $B'$ to be perpendicular to $\Sigma'$
so that $\Pi=\Span(\Sigma',B')$ and $\la B',w\ra=1$. This means $B'$ is
in the direction of the first column and we just need to rescale it by
$\lambda$. It
therefore is
\begin{equation}
B' = (\lambda^2-\ff12B^2)w + w^* + B,
\end{equation}
where $B$ is the $(p+q-2)$-dimensional vector with components
$(B_1,B_2,\ldots,B_{p+q-2})$. So, if we define $\alpha$ by
\begin{equation}
B' = \alpha w + w^* + B,
\end{equation}
then
\begin{equation}
  2\alpha = 2\lambda^2 - B^2.
\end{equation}
But we have already established $2\lambda^2=J^2$. So
\begin{equation}
 2\alpha = J^2 - B^2,
\end{equation}
as claimed (but unproven) in \cite{AM:K3p,me:lK3}.\footnote{
  An error in the original
  versions was corrected by E.~Diaconescu and was subsequently verified
  by other methods such as \cite{Ramgoolam:1998vc}.}

We then define $\Sigma$ as the projection of $\Sigma'$ in
$w^\perp/w$. From above, this is precisely the fiducial spacelike
$(p-1)$-plane in $\R^{p-1,q-1}$. $Z$ then acts on this $\Sigma_0$ to
yield an arbitrary $\Sigma$. We also have the rotation action of $Z$
on the $B$-field as expected via $\Int(m)$.

%%%%%%%%%%%%%%%%%%%%%%%%%%%%%%%%%%%%%%%%%%%%%%%%%%%%%%%%%% 

\section{Maximal $N_4=8$ Supergravity}

Toroidal compactification of supergravities in various dimensions
yields different numbers of supersymmetries in different
dimensions. Let $N_4$ denote the number of supersymmetries we would
have in 4 dimensions. Thus the maximal supergravity case is $N_4=8$.

One can take this maximal supergravity and find the moduli space
directly as in \cite{CJ:SO(8)}. It is instructive, however, to
proceed upwards from geometry as follows.

\subsection{From Geometry to Conformal Field Theory}

Begin with the type IIA or type IIB string in 10 dimensions, which we
know has maximal supersymmetry. We then compactify this on a torus to
preserve all the supersymmetries. We therefore expect to find the
moduli space of a flat metric on a torus. So we start with just this
moduli space as $\cM_0$.

A torus $(S^1)^n$, with a flat metric, is given by a lattice
$\Z^n\subset V $, where $V\cong\R^n$, with an inner product. So we can
choose a basis for this lattice and construct a matrix $A$ whose rows
are these basis vectors. We can assume the inner product is the usual
dot product. Any nondegenerate matrix will do and so $A\in\Gl(n,\R)$.

A rotation of these basis vectors will not change the torus. So we may
divide from the right (so we act within the rows of $A$) by
$\GO(n)$. Also, a lattice change of basis will not change the
underlying torus so we divide from left (to interchange the rows) by
$\GO(\Gamma)\cong\GO(n,\Z)$. The moduli space is thus
\begin{equation}
\bigslantt{\GO(\Gamma)}{\Gl(n,\R)}{\GO(n)}.
\end{equation}

We're shamelessly ignoring global issues so we can go ahead and write
this moduli space as
\begin{equation}
  \R_+\times\frac{\Sl(n,\R)}{\SO(n)},  \label{eq:torusM}
\end{equation}
where the $\R_+$ factor parametrizes the determinant, i.e., volume of
the torus.

We have a Cartan decomposition of the Lie algebra
\begin{equation}
  \frak{sl}(n,\C) = \frak{k}\oplus\frak{q},
\end{equation}
where $\frak{k}$ is antisymmetric matrices and $\frak{q}$ is traceless
symmetric matrices.  But $\frak{k}=\so(n)$ and so the tangent space of
the moduli space (\ref{eq:torusM}) is naturally given by symmetric
matrices. An infinitesimal change in the metric of a torus of fixed
volume corresponds to a traceless matrix. This naturally maps the
Killing form on $\frak{q}$ to the metric part of the Zamolodchikov
metric (\ref{eq:Zam}).

Now let us elevate this purely geometric compactification to a string
theory compactification. As well as a metric, we now also have a
$B$-field so we need to add the corresponding degrees of freedom to
the moduli space.

The $B$-field is a 2-form. That is, it lives in $\Lambda^2V$.  This
determines exactly how it transforms as a representation of
$\Sl(n,\R)$.

Finite-dimensional irreducible representations of a Lie algebra
$\frak{g}$ are classified by their highest weight, which can be
expressed as a nonnegative combination of fundamental weights
$\omega_i\in\frak{h}^*$. These fundamental weights are ``dual'' to the
simple roots $\alpha_j$ in the sense that
\begin{equation}
  \frac{2(\omega_i,\alpha_j)}{(\alpha_j,\alpha_j)} = \delta_{ij}.
\end{equation}
We can therefore naturally label an irreducible representation by
attaching nonnegative integers to the Dynkin diagram of $\frak{g}$. 

For example, the Dynkin diagram of $\sl(n,\R)$ is given by the
$A_{n-1}$ diagram with $n-1$ nodes.
\begin{equation}
  \dynkin[edge length=7mm,labels={\R,\R,\R,\R,\R}]{A}{ooo.oo}
\end{equation}
Every diagram in this section corresponds to split form and thus has
an $\R$ on each node. We will omit these labels for the rest of this
section.

If $\R^n$ is the fundamental
representation of $\sl(n,\R)$, then $\Lambda^k\R^n$, for $k< n$, is an
irreducible representation of $\sl(n,\R)$ given by (we omit
zeros)
\begin{equation}
  \begin{dynkinDiagram}[edge length=7mm,labels={,,,,,1}]{A}{oooo.oo.oo}
    \dynkinBrace*[\hbox{$k$ nodes}]68
  \end{dynkinDiagram}  \label{eq:Ank}
\end{equation}

As
such the $\Lambda^2 V$ representation of $\sl(n,\R)$ is given by:
\begin{equation}
  \begin{dynkinDiagram}[edge length=7mm,labels={,,,,,1}]{A}{ooo.oooo}
    %\dynkinBrace*[$n-1$]17
  \end{dynkinDiagram}  \label{eq:An1}
\end{equation}

We would like to form a corresponding Langlands decomposition that adds
the $B$-field to our torus. The only interpretation that is consistent
with the structure is
\begin{equation}
  \begin{split}
    \frak{m} &= \sl(n,\R)\\
    \frak{a} &= \R_+ \\
    \frak{n} &= \Lambda^2V.
  \end{split}
\end{equation}

Since $\frak{a}$ is rank one we need to add one node to the (restricted)
Dynkin diagram of $\frak{m}$ to build $\frak{g}$. The solution is to
add a node at the ``1'' label above:
\begin{equation}
  \begin{dynkinDiagram}[edge length=7mm,labels={,,,,,a}]{A}{ooo.oooo}
     \draw ($(root 6)+(-0.03mm,0.4mm)$) -- ++(0,0.7) ++(0,0.06) circle
     (0.05) node[anchor=west] {$\scriptstyle b$};
  \end{dynkinDiagram}  \label{eq:DnP}
\end{equation}

This is because the inner product between roots $a$ and $b$ above is
$-1$. So, upon projecting root $b$ into the root space of $\frak{m}$,
it is (up to a sign) the desired fundamental weight needed in
(\ref{eq:An1}).

The Dynkin diagram we obtain is $D_n$. The split form of this
corresponds to $\SO(n,n)$. So, the prediction is that we should use
the parabolic decomposition corresponding to removing the node $b$ in
(\ref{eq:DnP}).

What we are approaching is, of course, the Narain moduli space
\cite{N:torus} for a conformal field theory on a torus. But the point
is that we derive this purely by adding the $B$-field degree of
freedom to geometry.

Note that, at this point, we have implied that the
weights of $\frak{n}$ as a representation of $\frak{m}$ {\em
  include\/} the weights of $\Lambda^2 V$. There might be more. For
the case at hand we will see explicitly that this is not the
case. We will see this phenomenon of extra weights explicitly in an
example in the next section.

We can again use the analysis of section \ref{ss:K31} to write an
element of $\frak{g}$ as
\begin{equation}
  \begin{pmatrix}\mathsf{A}&\mathsf{B}\\\mathsf{C}&-\mathsf{A}^T\end{pmatrix},
\end{equation}
with $\mathsf{B}$ and $\mathsf{C}$ skew-symmetric.  The restricted roots are all the
roots $\pm L_i\pm L_j$, $i\neq j$ and the simple root labeled $b$ can
be taken to be $L_{n-1}+L_n$. So $\Pi'$ is given by the simple roots
$L_1-L_2,L_2-L_3,\ldots,L_{n-1}-L_n$. The Langlands decomposition then
corresponds to
\begin{itemize}
\item $\frak{m}$ is the algebra $\sl_n\C$ given by traceless
  matrices for the submatrix $\mathsf{A}$.
\item $\frak{a}$ is given by matrices $\mathsf{A}$ proportional to the
  identity.
\item $\frak{n}$ is given by the skew-symmetric submatrix $\mathsf{B}$.
\end{itemize}

So the $N\times A\times Z$ decomposition (\ref{eq:NAZ}) becomes
\begin{equation}
  \frac{\SO(n,n)}{\SO(n)\times\SO(n)} =
  \Lambda^2\R^n \times \R_+ \times \frac{\Sl(n,\R)}{\SO(n)}.  \label{eq:SOnnd}
\end{equation}

In the language of section \ref{s:para} we are forming the parabolic
subgroup using $R=H_1+H_2+\ldots+H_n$, and the $\alpha$'s generating
$\frak{n}$ are of the form $L_i+L_j$, $i<j$. So (\ref{eq:metricm=1}) is
\begin{equation}
  ds^2|_{m=1} = \frac{dB^2}{2\lambda^4} +
  \frac{n\,d\lambda^2}{\lambda^2}+dz,
     \label{eq:torusB}
\end{equation}
where $dB^2=\sum_{i<j}dB_{ij}^2$.

From our discussion at the start of this section, the volume of the
torus, $V$, corresponds to the determinant and so
\begin{equation}
  V = \lambda^n.
\end{equation}
Putting $y=\lambda^2$ we get
\begin{equation}
  ds^2|_{m=1} = \frac{dB^2 + \ff n2\,dy^2}{2y^2}.  \label{eq:torusB2}
\end{equation}

We want to identify this with the Zamolodchikov metric. As explained
in \cite{Moore:ZVol}, the Zamolidchikov metric for a CFT with a target
space a circle of radius $R$ is, up to an overall scale,
\begin{equation}
  ds^2 =
  \left(\frac{dR}R\right)^2.
\end{equation}
So, for a square torus with radii $R_i$ we can simply sum this over
$i$:
\begin{equation}
  ds^2 = \left(d\log\prod_{i=1}^n R_i\right)^2=(d\log V)^2,
\end{equation}
where $V$ is the volume.
Since the volume factor explicitly factorizes out above, this is true
for any shape of the torus. 

We are interested in the relative scaling of this volume contribution
and the $B$-field part. Again we can appeal to the form (\ref{eq:Zam}).
Thus the $B$-field metric must be warped by the volume factor in the
same way as the metric for deformations of the metric $g_{ij}$ is.

Focusing on changing the volume of the square torus, all the
$\delta g_{ii}$ terms will be equal and we will sum over $n$ of them.
Let $y=V^{\frac 2n}$ so that $y$ scales like areas within the torus,
and let $dB^2$ represent sum over all the changes in $dB_{ij}$ as
above.  Then we have a metric
\begin{equation}
  ds^2 = \frac{db^2+\ff n2dy^2}{y^2},
\end{equation}
for the $N$ and $A$ parts of the decomposition. This agrees with
(\ref{eq:torusB2}) up to an overall factor.

This is the same form as the hyperbolic plane for $n=2$. Indeed, this
is no coincidence. For the case of a 2-torus we have
\begin{equation}
  \begin{split}
    \frac{\SO(2,2)}{\SO(2)\times\SO(2)} &= \frac{\Sl(2,\R)}{\SO(2)}
                                         \times \frac{\Sl(2,\R)}{\SO(2)}\\
                                       &=
                                         \R\times\R_+ \times
                                         \frac{\Sl(2,\R)}{\SO(2)},
  \end{split}
\end{equation}
where the first line is from the equivalence
$\so(2,2)=\sl(2,\R)\oplus\sl(2,\R)$ and the second line is our
decomposition. So the $B$-field and volume are equivalent to the first
factor in the first line.

Exactly like the hyperbolic plane, we get the correct form for the $N$
part of the decomposition and thus we have established that
(\ref{eq:SOnnd}) is an equivalence of metric spaces. We have built the
Narain moduli space from geometry and the $B$-field.

\subsection{From Conformal Field Theory to String Theory}
\label{ss:cft2str}

Next let us go a step further and construct a full IIA or IIB string
compactification on a torus. This means we add two further degrees of
freedom --- the string coupling (or dilaton) and the Ramond-Ramond
moduli. The string coupling will obviously give us the $A=\R_+$ factor
of the decomposition.

The RR-moduli will form the $N$ factor and thus we need to know how
these form a representation of $\so(n,n)$. The $\so(n)\oplus\so(n)$
symmetry of the Narain model corresponds to rotations in the left and
right-moving sectors of the theory. The Ramond states are spinors
under such rotations and so the RR sector must be a product of two
spinors. This is enough information to determine that RR sector
scalars transform as spinors under $\so(n,n)$.

For example, consider the case $n=6$. The spinors of $\so(6)$ are
$\mathbf{4}$ and $\overline{\mathbf{4}}$. The subalgebra
$\so(6)\oplus\so(6)\subset\so(6,6)$ leads to decompositions of spinors
\begin{equation}
  \begin{split}
    \mathbf{32} &= (\mathbf{4},\mathbf{4})\oplus
                             (\overline{\mathbf{4}},\overline{\mathbf{4}})\\
    \mathbf{32}' &= (\mathbf{4},\overline{\mathbf{4}})\oplus
                             (\overline{\mathbf{4}},\mathbf{4})\\
   \end{split}
\end{equation}
corresponding to the type IIA and type IIB GSO projections.

The spinor representation of $\so(2n)$ is given by
\begin{equation}
  \begin{dynkinDiagram}[edge length=7mm,labels={,,,,,,1}]{A}{ooo.oooo}
     \draw ($(root 6)+(-0.03mm,0.4mm)$) -- ++(0,0.7) ++(0,0.06) circle
     (0.05);
  \end{dynkinDiagram}
\end{equation}
and so we add another node to the diagram at the labeled point to get
\begin{equation}
  \dynkin[edge length=8mm,backwards]{E}{ooooo.ooo}
\end{equation}

There are now $n+1$ nodes in the diagram.  So the prediction is that
we produce the group $E_{n+1}$ and its corresponding symmetric
spaces. As always in this section, we are in the {\em split form\/} of
the group so we get the symmetric space corresponding to
$E_{n+1(n+1)}$ as listed in \cite{Helgason:Lie}.

For example, the
$n=6$ case yields
\begin{equation}
  \frac{E_{7(7)}}{\SU(8)} = (\R\oplus\R^{32}) \times
    \R_+ \times \frac{\SO(6,6)}{\SO(6)\times\SO(6)}.  \label{eq:E77toNar}
\end{equation}
Note that the representation theory of this parabolic decomposition
yields a singlet $\R$ in addition to the spinor $\R^{32}$. This should
not come as a surprise as our highest weight argument above using
Dynkin diagrams only assured that the $\frak{n}$ factor was a
representation of $\frak{m}$ with the given highest weight. There was no
guarantee that $\frak{n}$ should be irreducible and in general it is
not.

We have imposed the correct $\Int(m)^*$ action to find
(\ref{eq:E77toNar}) but what about the $\Int(a)^*$ action? We show in
the appendix that, if $(R,R)=2$ then $\alpha(R)=1$ for the
$\mathbf{32}$ spinor and $\alpha(R)=2$ for the singlet. So we have a
metric
\begin{equation}
  ds^2|_{m=1} = \frac{d\theta^2}{2\lambda^4} + \frac{\sum
    dS_a^2}{2\lambda^2} + \frac{2d\lambda^2}{\lambda^2},
\end{equation}
where $\phi$ is our singlet and $S_a$ are the components of the spinor.
Putting $\phi=\lambda^2$ and rescaling overall by 2 we thus obtain
\begin{equation}
  ds^2|_{m=1} = \frac{d\phi^2+d\theta^2}{\phi^2} + \frac{\sum
    dS_a^2}{\phi}.
\end{equation}
The first factor we recognize as the hyperbolic plane again this time
parametrized by the dilaton-axion as expected. The RR fields $S_a$
have a metric which is half as warped as the dilaton factor. This
resonates with Shenker's observation that the strength of the
nonperturbative effects coming from the RR sector has a power which is
``half'' that of the dilaton \cite{Shenker:1990uf}.

We can perform a very similar computation for other values of $n$. 
Of course, we only have the conventional exceptional groups
$E_6,E_7,E_8$ when $5\leq n\leq 7$. It is easy to see that $D_5$ plays
the role of ``$E_5$'' and $A_4$ plays the role of ``$E_4$''. The case
of ``$E_3$'' corresponds to $A_1\oplus A_2$. This is because $n=2$
corresponds to $\so(2,2)$ whose Dynkin diagram is two unconnected
nodes. The RR-fields contribute a spinor $(\mathbf{2},\mathbf{1})$
which attaches a node to one of these two:

\begin{equation}
  \begin{tikzpicture}[scale=1.0,
          baseline={([yshift=-.8ex]current bounding box.center)}]
    \path (0,0) node[circle,draw,inner sep=1pt] (A) {}
    (0,0.7) node[circle,draw,inner sep=1pt] (B) {};
  \end{tikzpicture}\qquad\to\qquad
  \begin{tikzpicture}[scale=1.0,
          baseline={([yshift=-.8ex]current bounding box.center)}]
    \path (0,0) node[circle,draw,inner sep=1pt] (A) {}
    (0,0.7) node[circle,draw,inner sep=1pt] (B) {}
    (-0.7,0) node[circle,draw,inner sep=1pt] (C) {};
    \draw (A) -- (C);
  \end{tikzpicture}
\end{equation}

\vspace{3mm}

For compactification on a circle, i.e., $n=1$, the Lie algebra
$\so(1,1)$ is no longer semi-simple and the Dynkin diagram picture
breaks down completely. 

In the other direction, for $n=8$ we have $E_9$ and here the moduli
space picture using finite-dimensional symmetric spaces breaks down,
as is well-known for two-dimensional theories.

\subsection{From Geometry to M-Theory}

Rather than the two-step process of using string theory, going from
$A_{n-1}$ to $D_n$ to $E_{n+1}$ that we did above, we could instead go
directly from $A_n$ to $E_{n+1}$ by adding a single node. So this goes
from the torus geometry directly to the complete supersymmetric theory.

This differs from string theory in two ways:
\begin{enumerate}
\item For $A_n$ we have the geometry of an $(n+1)$-dimensional
  torus. So our primary theory being compactified must have been in
  11 dimensions rather than 10.
\item Since the node we are adding is third from the right, from
  (\ref{eq:Ank}) we are using a 3-form.
\end{enumerate}
So, purely from the combinatorics of the Dynkin diagram, we deduce the
existence of an 11-dimensional theory with a 3-form fundamental degree
of freedom. We might claim we have predicted M-theory purely from a
representation theory argument.

Let us give an example of the M-theory moduli space equivalence. In 4
dimensions we add a node to $A_7$ to get $E_8$. This yields:
\begin{equation}
  \frac{E_{7(7)}}{\SU(8)} = (\R^{35}\oplus\R^7) \times \R_+
  \times \frac{\Sl(7,\R)}{\SO(7)}.
\end{equation}
The $\R_+$ factor is the volume of the 7-torus and the final factor is
its shape. The $\R^{35}$ term is $\Lambda^3\R^7$ and is the
compactification of the 3-form on the 7-torus. However, in 4
dimensions we also have that 2-form fields are dual to scalars. We can
thus compactify the M-theory 3-form on a one-cycle of the torus and
this yields the extra $\R^7$ term in $\frak{n}$.

\subsection{Compactifying on a circle}

We have not considered all the possibilities of adding nodes to the Dynkin
diagram. The remaining possibility in each case is to add a node on
the left end of the Dynkin diagram string. This is equivalent to
replacing $n$ by $n+1$ in each case. Thus we are compactifying down
one dimension and the only one-dimensional compact manifold is a
circle.

So we begin with a supergravity theory with moduli space given by $Z$
in our decomposition. The $A=\R_+$ factor is the radius of the circle
on which we compactify. This is like a Kaluza--Klein
compactification. The $\frak{n}$ must correspond to the extra moduli
that arise from Wilson lines of the starting gauge group.

As an example let us compactify from 4 down to 3 dimensions.
This is the Dynkin diagram manipulation
\begin{equation}
  \dynkin[edge length=6mm,backwards,labels={,,,,,,1}]E7  \qquad\to\qquad
  \dynkin[edge length=6mm,backwards]E8  \label{eq:E7toE8}
\end{equation}

This corresponds to
\begin{equation}
  \frac{E_{8(8)}}{\SO(16)} = (\R^{56}\oplus\R) \times \R_+ \times
  \frac{E_{7(7)}}{\SU(8)}.
\end{equation}
The irreducible representation corresponding to the left diagram in
(\ref{eq:E7toE8}) is the $\mathbf{56}$ of $E_7$ so we predict this is
how the gauge fields of this supergravity theory transform. This was
derived explicitly in \cite{CJ:SO(8)} with some effort. Note that in 3
dimensions the circle itself generates a $\GU(1)$ gauge symmetry and
the associated 1-form is dual to a scalar. This yields the extra $\R$
appearing in $\frak{n}$.

\subsection{Limits in the Moduli Space}

The more conventional use of our decompositions via parabolic
subgroups is to go in the other direction than the above. We saw in
section \ref{ss:limits} that removing nodes from the Dynkin diagram to
find limits is quite general.

We use the $A=\R_+$ factor in the moduli space to go to some limit
when removing a single node. Since the metric on this factor is
universally $ds^2=da^2/a^2$, this limit is always infinite distance.

This is a generalization of the ``Group Disintegrations'' of Julia
\cite{Jul:dis}. For above we understand decompactification of a
circle, going to a large M-theory torus or going to a weak
string-coupling limit as follows:
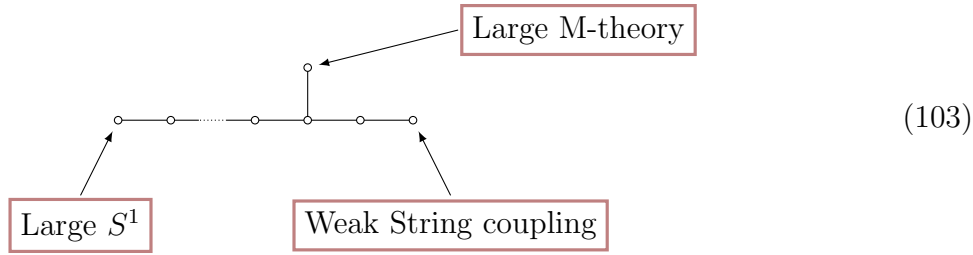
\begin{equation}
  \begin{dynkinDiagram}[edge length=7mm,backwards]{E}{ooooo.oo}
    \draw[latex-] (root 1) -- ++(-5mm,-10mm) node[rectangle, very thick,
    draw=red!50!black!50,yshift=-4mm]
            {Weak String coupling};
     \draw[latex-] (root 7) -- ++(5mm,-10mm) node[rectangle, very thick,
    draw=red!50!black!50,yshift=-4mm] {Large $S^1$};
     \draw[latex-] (root 2) -- ++(-20mm,5mm) node[rectangle, very thick,
    draw=red!50!black!50,xshift=16mm,yshift=0mm] {Large M-theory};
  \end{dynkinDiagram}
\end{equation}
  
There are other possibilities, however. For example, consider the
following node deletion
\begin{equation}
  \dynkin[edge length=7mm,backwards]E{ooooooxo}
\end{equation}
This yields
\begin{equation}
  \frac{E_{8(8)}}{\SO(16)} = \R^{83} \times \R_+
    \times \left(\frac{E_{6(6)}}{\Sp(4)}\times
      \frac{\Sl(2,\R)}{\GU(1)}\right),
\end{equation}
where the $\frak{n}=\R^{83}$ decomposes as a representation of
$E_{6(6)}\times \Sl(2,\R)$ as
\begin{equation}
  (\mathbf{27},\mathbf{2})\oplus(\mathbf{27},\mathbf{1})\oplus
  (\mathbf{1},\mathbf{2}).
\end{equation}
The obvious interpretation of this is a large 2-torus
decompactification of 3-dimensional supergravity to 5-dimensional
supergravity. The $\Sl(2,\R)/\GU(1)$ is the modulus of the torus while
the $\R_+$ is the area. The gauge fields in 5-dimensional supergravity,
which live in a $\mathbf{27}$ of $E_6(6)$, can give scalars a la
Kaluza-Klein by wrapping either of the cycles of the 2-torus. This
accounts for the highest-weight $(\mathbf{27},\mathbf{2})$ in
$\frak{n}$. The other parts of $\frak{n}$ come from peculiarities of
3-dimensions where vectors are dual to scalars.

%%%%%%%%%%%%%%%%%%%%%%%%%%%%%%%%%%%%%%%%%%%%%%%%%%%

\section{$N_4=6$ Supergravity}

Let us now analyze the Dynkin diagram combinatorics of lower
supergravities in the more conventional direction using
disintegrations. That is, we proceed from lower to higher dimensions.
It is well-known that $N_4=7$ supersymmetry automatically promotes
itself to $N_4=8$ \cite{CJ:SO(8)}. So the next case of interest is
$N_4=6$. Only in the maximal $N_4=8$ case do we deal with split forms
of Lie groups. So from now, on, the restricted roots are no longer the
same as the usual roots.

Let us begin with the maximal Dynkin diagram, i.e., the analogue of
$E_8$ in the previous section. This corresponds to supergravity in
three-dimensions and it is known \cite{Jul:dis} that this has a moduli
space
\begin{equation}
  \frac{E_{7(-5)}}{\SO(12)\times\SU(2)}.
\end{equation}

The restricted root system of $E_{7(-5)}$ (also known as {\bf E
  VI}), corresponds to $F_4$ \cite{Knapp:beyond}.
The restricted Dynkin diagram for $E_{7(-5)}$ is then the $F_4$ Dynkin
diagram with $\frak{m}_\alpha$'s given by $\sl(2,\R)$ or $\sl(2,\H)\cong\so(1,5)$:
\begin{equation}
  \dynkin[edge length=8mm,labels={\R,\R,\H,\H}]{F}{oooo}  \label{eq:F4}
\end{equation}
Note that this diagram is maximal in the sense we cannot add any
further nodes. This fits in with the general idea that we cannot
compactify below 3 dimensions as we saw with $E_8$ in the previous
section.

We now have two ``ends'' from which we may pull nodes for parabolic
subgroups. These two possibilities will correspond to circle
decompactifications and weak string coupling limits in a sense we make
precise later. Note we have lost any obvious M-theory interpretations
as we have no ``third end''.

The parabolic subgroup
\begin{equation}
\dynkin[edge length=8mm,labels={\R,\R,\H,\H}]{F}{xooo}
\end{equation}
yields
\begin{equation}
  \frac{E_{7(-5)}}{\SO(12)\times\SU(2)}=
  (\R^{32}\oplus\R)\times\R_+\times\frac{\SO^*(12)}{\GU(6)},
\end{equation}
which is decompactification to 4 dimensions.

The parabolic subgroup
\begin{equation}
\dynkin[edge
length=8mm,labels={\R,\R,\H,\H}]{F}{ooox}
\end{equation}
yields
\begin{equation}
  \frac{E_{7(-5)}}{\SO(12)\times\SU(2)}=(\R^{16}\oplus\R^{16}
\oplus\R^{10})\times\R_+\times\frac{\SO(3,7)}{\SO(3)\times\SO(7)}. \label{eq:AO37}
\end{equation}
We claim this is a weak string coupling limit and interpret it shortly. We
can also go to the weak string coupling of the 4-dimensional theory by
using
\begin{equation}
\dynkin[edge
length=8mm,labels={\H,\H,\R},backwards]{C}{xoo}
\end{equation}
to get\footnote{Note the equivalence $\SO^*(8)\cong\SO(2,6)$ is
  useful here.}
\begin{equation}
  \frac{\SO^*(12)}{\GU(6)}=(\R^{8}\oplus\R^{8}
     \oplus\R)\times\R_+\times\frac{\SO(2,6)}{\SO(2)\times\SO(6)}.
\end{equation}
In the $N_4=8$ examples in the previous section, the weak string
coupling limits all yielded Narain moduli spaces. The rightmost component
here looks like a Narain moduli space for a 6-torus but 4 of the 6
left-movers have disappeared. Actually, we know exactly how to get
this. It is an {\em asymmetric orbifold\/} as described in
by Dabholkar and Harvey \cite{Dabholkar:1998kv}.

The Narain moduli space may be viewed as the Grassmannian of
space-like 6-planes $\Pi$ relative to a lattice
$\Gamma\subset\R^{6,6}$. Asymmetric orbifolds use chiral symmetries
that only appear at specific points in the moduli space. In this case,
we need a timelike $D_4$ lattice to lie perpendicular to $\Pi$. This
restricts the Grassmannian down to $\SO(2,6)/(\SO(2)\times\SO(6))$ as
we observe. We refer to \cite{Dabholkar:1998kv} for details of the
construction. Similarly this also explains the
$\SO(3,7)/(\SO(3)\times\SO(7))$ in three dimensions.

If we paid careful attention to the global structure of the moduli
space we should, presumably, be able to discover precisely the
$D_4$-lattice of \cite{Dabholkar:1998kv}. While this would be
interesting to do, we claim we already know it must work since this
asymmetric orbifold model gives an $N_4=6$ theory.

So what's interesting here is that the Dabholkar--Harvey asymmetric
orbifold appears not so much to give {\em a\/} construction of an
$N=6$ theory in four dimensions as {\em the\/} construction (assuming
a weakly coupled string theory limit). It accounts for the full moduli
space of possibilities. All the other known methods of construction
should be equivalent to this.\footnote{There is another asymmetric
  orbifold construction of an $N_4=8$ in \cite{Dabholkar:1998kv} that
  forces the spacelike 6-plane to be perpendicular to a 6-dimensional
  lattice. This freezes all the Narain moduli out but there are twist
  fields which yield additional scalars and thus moduli.} The Hodge
diamond for the $N=(2,2)$ superconformal field theory giving the right
spectrum for this must be
\begin{equation}
  \setlength\arraycolsep{2pt}
  \begin{eqmatrix}[10pt]
    &&&1&&&\\
    &&1&&3&&\\
    &1&&3&&3&\\
    1&&3&&3&&1\\
    &3&&3&&1&\\
    &&3&&1&&\\
    &&&1&&&\\
\end{eqmatrix}  \label{eq:N=6HD}
\end{equation}
and this is consistent with the spectra found by other methods, e.g.,
\cite{Kreuzer:1993uy}.

Meanwhile, the other deletion
\begin{equation}
\dynkin[edge
length=8mm,labels={\H,\H,\R},backwards]{C}{oox}
\end{equation}
gives
\begin{equation}
  \frac{\SO^*(12)}{\GU(6)}=\Lambda^2\R^{6}\times\R_+\times\frac{\Sl(3,\H)}{\Sp(3)},
\end{equation}
which is decompactification to $N=3$ in 5 dimensions.

We can then go to the weak string-coupling limit of this 5-dimensional
theory:
\begin{equation}
\dynkin[edge
length=8mm,labels={\H,\H},backwards]{A}{xo} \label{eq:F4HH}
\end{equation}
gives\begin{equation}
  \frac{\Sl(3,\H)}{\Sp(3)}=(\R^{4}\oplus\R^4)\times\R_+\times\frac{\SO(5,1)}{\SO(5)},
\end{equation}
which is the Narain moduli space of the asymmetric orbifold again.

Because of the symmetry of the Dynkin diagram in (\ref{eq:F4HH}),
pulling the left node off is the same.  We might therefore be tempted
to view the same decomposition as a large radius limit taking us to a
6-dimensional theory. Such a theory would have $N=(2,1)$ supersymmetry
in 6 dimensions and the associated $N=(2,2)$ superconformal field
theory would have a Hodge diamond
\begin{equation}
  \setlength\arraycolsep{2pt}
  \begin{eqmatrix}[10pt]
    &&&1&&&\\
    &&0&&2&&\\
    &1&&0&&1&\\
    &&2&&0&&\\
    &&&1&&&\\
\end{eqmatrix}  \label{eq:N=21HD}
\end{equation}
The Hodge diamond in (\ref{eq:N=6HD}) is given by the product of this
Hodge diamond with that of a 2-torus, which seems promising. However,
the Hodge numbers of (\ref{eq:N=21HD}) are not consistent with a
modular-invariant elliptic genus \cite{Eg:K3}, and, indeed, $N=(2,1)$
supergravity appears to be anomalous. So this decompactification to 6
dimensions is fictitious. The only limit is to a weakly-coupled string.

We have therefore exhaustively arrived at the following.
\begin{prop}
  For the $N_4=6$ family of supergravities, the maximal restricted
  Dynkin diagram is $F_4$ as in (\ref{eq:F4}). Pulling a root labeled
  $\R$ off the end of the Dynkin diagram corresponds to a
  decompactification, while pulling a root labeled $\H$ off the end
  corresponds to a weak string coupling limit.
\end{prop}

%%%%%%%%%%%%%%%%%%%%%%%%%%%%%%%%%%%%%%%%%%%%%%%%%%%%%%%%%%%%%%%%%%%

\section{$N_4=5$}

A three dimensional theory with $N=10$ supersymmetry has a moduli
space $E_6(-14)/((\SO(10)\times\GU(1))$. The numerator, also known as
{\bf E III}, has restricted Dynkin diagram $(BC)_2$
\cite{Knapp:beyond}:
\begin{equation}
  \dynkin[edge
length=10mm,arrows=false,labels={{\so(1,7)},{\su(1,5)}}]{C}{oO}
\end{equation}
We have root spaces generated by simple roots corresponding to
$L_1-L_2$, $L_1$ and $2L_1$ with multiplicities 6, 8 and 1
respectively.

We can find a weak string coupling limit:
\begin{equation}
  \begin{split}
  &\qquad\dynkin[edge
length=10mm,arrows=false,labels={{\so(1,7)},{\su(1,5)}}]{C}{ox}\\
\frac{E_{6(-14)}}{\SO(10)\times\GU(1)} = (\R^8&+\R^8+\R^8) \times \R_+\times
   \frac{\SO(1,7)}{\SO(1)\times\SO(7)},
  \end{split}
\end{equation}
where $N$ has 2 spinors and 1 vector (one of each kind). So we find
the restricted Narain moduli space for a 7-torus suggesting we have an
asymmetric orbifold. Indeed, \cite{Dabholkar:1998kv} construct an
asymmetric orbifold based on a $A_6$-lattice being orthogonal to the
Narain plane with the correct supersymmetry.

We can decompactify to 4 dimensions:
\begin{equation}
  \begin{split}
  &\qquad\dynkin[edge
length=10mm,arrows=false,labels={{\so(1,7)},{\su(1,5)}}]{C}{xO}\\
\frac{E_{6(-14)}}{\SO(10)\times\GU(1)} = (\R^{20}&+\R) \times \R_+\times
   \frac{\SU(1,5)}{\textrm{S}(\GU(1)\times\GU(5))},
  \end{split}
\end{equation}

This four-dimensional supergravity theory is given by a
compactification on a {\em rigid\/} superconformal field theory. Indeed,
the asymmetric orbifold construction above has the $A_6$ lattice
filling all the timelike directions for the Narain moduli space. The
Hodge diamond of this rigid theory is
\begin{equation}
  \setlength\arraycolsep{2pt}
  \begin{eqmatrix}[10pt]
    &&&1&&&\\
    &&0&&3&&\\
    &0&&0&&3&\\
    1&&0&&0&&1\\
    &3&&0&&0&\\
    &&3&&0&&\\
    &&&1&&&\\
\end{eqmatrix}  \label{eq:N=5HD}
\end{equation}
The same theory was constructed in \cite{Kreuzer:1993uy} as an
orbifold of a Gepner model.  We can do a decomposition of the moduli
space for the weak string-coupling limit by deleting the reduced
Dynkin diagram completely:
\begin{equation}
  \begin{split}
  &\qquad\dynkin[labels={{\su(1;5)}}]{A}{x}\\
\frac{\SU(1,5)}{\textrm{S}(\GU(1)\times\GU(5))} &= (\R^{8}+\R) \times \R_+.
  \end{split}
\end{equation}
So $Z$ is trivial as we would expect for a rigid theory. The axion and
8 RR moduli give the $\frak{n}$ factor.

This family of theories cannot be decompactified to 5 dimensions.

%%%%%%%%%%%%%%%%%%%%%%%%%%%%%%%%%%%%%%%%%%%%%%%%%%%%%%%%%%%%%%%%%%%

\section{$N_4=4$}
Once we reach $N_4=4$ there are other possibilities beside the
supergravity multiplet and so the moduli space is not completely
determined. $N=8$ supergravity in 3 dimensions has a moduli space
$\SO(8,m)/(\SO(8)\times \SO(m)$ as examined in \cite{Marcus:1983hb},
for example. The integer $m$ depends on the number of vector
supermultiplets which, unlike the above examples, are independent of
the supergravity multiplet.

$N_4=4$ models can also be created by compactification of the
heterotic string. This leads to possible multiple interpretations of the
parabolic decompositions. For example, the decomposition we considered
earlier (\ref{eq:K3a}) as the geometric limit of a K3 conformal field
theory could equally well be viewed as decompactifying a circle in
the Narain picture of the heterotic string on a 4-torus.

Indeed, there is a plethora of possibilities of producing models in
various dimensions with $N_4=4$ coming from the heterotic string,
M-theory and F-theory as described extensively in
\cite{mndog:triples}, for example.

Suppose we first assume that there is a direct type II interpretation.  A
$c=6$ superconformal field that can be used for a type II string
compactification has its spectrum tightly constrained by modular
invariance of the elliptic genus as we stated above. Relatedly,
supergravity theories in 6 dimensions can be anomalous
\cite{Sei:K3}. Essentially, the only two allowed Hodge diamonds
correspond to a 4-torus, in which case we are back to maximal
supersymmetry, or a K3 surface with $h^{1,1}=20$. This latter case
corresponds to $m=24$.

To avoid $m=24$ in the type II context we must be in a situation where
we cannot decompactify to 6 dimensions. One way is another asymmetric
orbifold of \cite{Dabholkar:1998kv} that has no vector
supermultiplets. This corresponds to $m=2$ with restricted Dynkin
diagram $B_2$:
\begin{equation}
  \dynkin[edge
length=8mm,labels={\R,{\so(1,7)}}]{B}{oo}
\end{equation}
Deletion of the left node is
\begin{equation}
  \begin{split}
  &\qquad\dynkin[edge length=8mm,labels={\R,{\so(1,7)}}]{B}{xo}\\
\frac{\SO(2,8)}{\SO(2)\times\SO(8)} &= \R^{8} \times \R_+
 \times \frac{\SO(1,7)}{\SO(7)}
  \end{split}
\end{equation}
This is the weak-coupling limit of the 3-dimensional theory. The
latter factor is the conformal field moduli space of the asymmetric
orbifold with a timelike $E_6$ lattice restricting the Narain
spacelike 7-plane \cite{Dabholkar:1998kv}.

The right node gives decompactification to 4 dimensions:
\begin{equation}
  \begin{split}
  &\qquad\dynkin[edge length=8mm,labels={\R,{\so(1,7)}}]{B}{ox}\\
\frac{\SO(2,8)}{\SO(2)\times\SO(8)} &= ((\R^{2})^6+\R) \times \R_+
 \times \frac{\Sl(2,\R)}{\GU(1)}
  \end{split}
\end{equation}
The last factor is the dilaton--axion system of the pure supergravity
in 4 dimensions. Importantly, we cannot decompactify to 6 dimensions.

For the K3-related family, we have the 3-dimensional $N=8$
supergravity with $m=24$ given by
\begin{equation}
\dynkin[edge
length=8mm,labels={\R,\R,\R,\R,\R,\R,\R,{\so(1,17)}}]{B}{oooooooo}. \label{eq:m24}
\end{equation}
Let us search for all the interpretations we expect for the many
limits of this theory.

Knocking off the left node
\begin{equation}
\dynkin[edge
length=8mm,labels={\R,\R,\R,\R,\R,\R,\R,{\so(1,17)}}]{B}{xooooooo},
\end{equation}
we find
\begin{equation}
\frac{\SO(8,24)}{\SO(8)\times\SO(24)}=
\R^{30}\times
\R_+\times \frac{\SO(7,23)}{\SO(7)\times\SO(23)}.
\end{equation}
This has an interpretation as the weak heterotic string coupling limit exhibiting the Narain
moduli space of the 7-torus. The $\R^{30}$ factor comes from the
$\GU(1)^{30}$ gauge bosons which are dual to scalars in 3 dimensions.

We can find the M-theory interpretation of this 3-dimensional model by
killing 2 nodes:
\begin{equation}
\begin{split}&\dynkin[edge
 length=8mm,labels={\R,\R,\R,\R,\R,\R,\R,{\so(1,17)}}]{B}{xoooxooo},\\
\frac{\SO(8,24)}{\SO(8)\times\SO(24)}&=
(\R^{88}+\R^4+\R^{22}+\R^6+\R^4)\times
\R_+^2\times \left(\frac{\Sl(4)}{\SO(4)}\times\frac{\SO(3,19)}{\SO(3)\times\SO(19)}
\right).
\end{split}  \label{eq:MN4}
\end{equation}
We see directly the classical moduli space of $T^4$ and the K3.
Now $A$ is $\R_+^2$ since we need volumes for both the torus and the K3.
The $\R^{88}$ factor is the manifest $(\mathbf{4},\mathbf{22})$
representation from the rightmost cross as 3-forms created from
1-forms on $T^4$ and 2-forms on the K3. The $\R^4$ factor are the
3-forms on $T^4$. The rest comes from gauge bosons dualized to scalars.

We decompactify to 4 dimensions by deleting second from the end:
\begin{equation}
  \begin{split}
    &\dynkin[edge
length=8mm,labels={\R,\R,\R,\R,\R,\R,\R,{\so(1,17)}}]{B}{oxoooooo}\\
\frac{\SO(8,24)}{\SO(8)\times\SO(24)}&=
(\R^{28\times2}+\R)
   \times\R_+\times\left(\frac{\SO(6,22)}{\SO(6)\times\SO(22)}
                                       \times\frac{\Sl(2,\R)}{\GU(1)}\right).
  \end{split}
\end{equation}

We won't go through all the possibilities for higher dimensions here
as they are straight-forward. One recurring possibility is the
following deletion for a diagram with $n$ nodes, with $n<8$,
\begin{equation}
\dynkin[edge
length=8mm,labels={\R,\R,\R,{\so(1,17)}}]{B}{oo.ox}  \label{eq:so117R}
\end{equation}
This yields
\begin{equation}
  \frac{\SO(n,n+16)}{\SO(n)\times\SO(n+16)} =
  ((\R^n)^{16} + \Lambda^2\R^n) \times \R_+ \times
  \frac{\Sl(n,\R)}{\SO(n)}.
\end{equation}
This is the geometric interpretation of the moduli space of conformal
theories as a heterotic compactification on a torus. The $(\R^n)^{16}$
factor is the Wilson lines of the heterotic gauge group part, the
$\Lambda^2\R^n$ is the $B$-field and the last factor is, of course,
the shape of the torus.

That said, what if we do the right-most node removal (\ref{eq:so117R})
in the $n=8$ case of our original 3-dimensional theory (\ref{eq:m24})?
This is the full string theory moduli space, not just the CFT, so we can't
use the heterotic interpretation above. The right factor is telling us
that we have the geometry of an 8-torus so, to get to 3 dimensions we
must be compactifying an 11-dimensional theory. But it's not
M-theory. There is a 3-form in M-theory and the first factor of the
decomposition sees no 3-form. Indeed, we already saw the
M-theory compactification appearing perfectly in (\ref{eq:MN4}).

What this limit corresponds to is not clear to us. Since we have the
manifest moduli space of an 8-torus we are free to go to
decompactification limits and thus produce higher-dimensional
theories. Indeed, it's not clear why we can't go all the way to 11 dimensions.

Another possibility for avoiding $m=24$ in the 3-dimensional theory is
that there may be {\em no\/} obvious type II interpretation. Thus we
avoid any anomaly constraint above while being able to decompactify
above dimension 6. As mentioned above, this leads to many
possibilities the oldest of which is the CHL construction
\cite{CHL:bigN,CP:ao}. Quotienting by a $\Z_2$ symmetry that switches
the two $E_8$'s of the heterotic string combined with a particular
translation in the torus directions yields $m=16$ vector
supermultiplets where compactifying to 3 dimensions. So this CHL
family is based on the 3-dimensional diagram
\begin{equation}
\dynkin[edge
length=8mm,labels={\R,\R,\R,\R,\R,\R,\R,{\so(1,9)}}]{B}{oooooooo}.
\end{equation}

Pulling the left node off gives us the weakly-coupled heterotic
string limit while pulling the next-to-left node off gives us a
decompactification to 4 dimensions as above. There is no node removal that
would {\em require\/} a weakly coupled type II string interpretation. This is
not to say that there might be some interesting novel type II interpretation
of CHL compactifications that has yet to be determined, but we see
that it is not compulsory to have one.

Pulling off the right node gives
\begin{equation}
  \frac{\SO(8,16)}{\SO(8)\times\SO(16)} =
  ((\R^8)^{8} + \Lambda^2\R^8) \times \R_+ \times
  \frac{\Sl(8,\R)}{\SO(8)},
\end{equation}
giving us another peculiar 8-torus picture like we had above.

Note we can decompactify the CHL string almost as much as we want but
we know something goes wrong if we try to decompactify all the way to
ten dimensions since we needed the translation around a torus. This is
reflected in the discrete group quotient of the moduli space, which we
are not considering here.

\section{Discussion}

We have a seen a few applications of using parabolic subgroups to
decompose symmetric spaces. It is important to note that all of our
analysis has been in terms of the ``Teichmuller'' symmetric space
$G/K$. Of course, in the context of string theory, it is key to take
into account dualities and thus obtain a true moduli space
$\Gamma\backslash G/K$, where $\Gamma$ is a discrete subgroup of $G$.

A more careful analysis taking $\Gamma$ into account shows that
remnants of $\Gamma$ remain in the $N$, $A$ and $Z$ components. The
$Z$ part is the duality group or modular group of the subtheory while
the $N$ part results in the toriodal structure. This latter part is
the familiar periodicity in the $B$-field or $RR$-fields, for
example. What is perhaps more interesting is the part of $\Gamma$ that
is ``lost'' and has no simple interpretation in terms of $N$, $A$ or
$Z$. This is, of course, understood in many examples, but it would be
interesting to find systematics in this process. In many respects,
this is an old mathematical problem related to the motivation of
\cite{Borel:decomp} in the first place and progress might be difficult.

%%%%%%%%%%%%%%%%%%%%%%%%%%%%%%%%%%%%%%%%%%%%%%%%%%%%%%%%%%%%%%%%%%%
\section*{Acknowledgments}

I thank David Morrison for the original collaboration that made me
aware of this method. I also thank Ilarion Melnikov and Ronen Plesser
for helpful discussions.

\appendix

\section{$E_{n+1}$ computations}

Let us compute the details of the parabolic decomposition metrics for
the case $\frak{g}=\frak{e}_7$. It is easy to generalize this to the
other $\frak{e}_{n+1}$ cases. Following the conventions of
\cite{Knapp:beyond}, we enumerate the simple roots, $\alpha_i$, from the
following Dynkin diagram:
\begin{equation}
  \dynkin[edge length=8mm,backwards,labels={1,2,3,4,5,6,7}]E{ooooooo}
\end{equation}
The conventional way to describe $\frak{h}^*$ is by embedding
$\frak{e}_7$ into $\frak{e}_8$. So we take $e_1,e_2,\ldots,e_8$ to be
an orthonormal basis of $\R^8$ but then we restrict to the orthogonal
complement of $e_7+e_8$ to obtain (the real slice of)
$\frak{h}^*$. The simple roots are then
\begin{equation}
  \begin{split}
    \alpha_1&=\ff12(e_8-e_7-e_6-e_5-e_4-e_3-e_2+e_1)\\
    \alpha_2&=e_2+e_1\\
    \alpha_3&=e_2-e_1\\
    \alpha_4&=e_3-e_2\\
    \alpha_5&=e_4-e_3\\
    \alpha_6&=e_5-e_4\\
    \alpha_7&=e_6-e_5.
  \end{split}
\end{equation}
Then define $H_i\in\frak{h}$ by $e_i(H_j)=\delta_{ij}$. So any
element $\sum a_iH_i\in\frak{h}$ must satisfy $a_7+a_8=0$.

Consider first removing node 1 corresponding to the passage between
conformal field theory and string theory in section
\ref{ss:cft2str}. The subalgebra $\frak{a}\subset\frak{h}$ has a single generator
which we call $R$. The $\alpha$'s in (\ref{eq:akerB}) are given by the
remaining roots which determines (up to a scale)
\begin{equation}
  R = H_8-H_7.
\end{equation}
Now we take all the positive roots of $\frak{e}_7$ and compute
$\alpha(R)$. Those with $\alpha(R)=0$ contribute towards $\frak{m}$,
and those with $\alpha(R)>0$ give $\frak{n}$. The Lie algebra
brackets within $\frak{g}$ then gives $\frak{n}$ the structure of a
representation of $\frak{m}\oplus\frak{a}$. The 63 positive roots of
$\frak{e}_7$ divide as 
\begin{itemize}
\item $e_i\pm e_j$ with $j<i\leq6$ have $\alpha(R)=0$ and contribute
  the positive roots for $\so(12)$.
  \item $\ff12(+e_8-e_7\pm e_6\pm e_5\pm e_4\pm e_3\pm e_2\pm e_1)$
    with an even number of $+$ signs have $\alpha(R)=1$ and generate a
    spinor $\mathbf{32}$ of $\so(12)$.
  \item $e_8-e_7$ has $\alpha(R)=2$ and is an invariant ${\mathbf{1}}$ under
    $\so(12)$.
\end{itemize}

For the M-theory interpretation we remove node 2. The remaining roots
all annihilate $R=H_1+H_2+H_3+H_4+H_5+H_6-2H_7+2H_8$. The 63 positive
roots of $E_7$ then split as
\begin{itemize}
\item 21 with $\alpha(R)=0$ to contribute towards building
  $\sl(7)$.
\item $\Lambda^3\R^7=\mathbf{35}$ with $\alpha(R)=2$ and
\item $\mathbf{7}$ with $\alpha(R)=4$.
\end{itemize}

For the circle decompactification we remove node 7. The remaining
roots all annihilate $R=2H_6-H_7+H_8$. 
The 63 positive roots of $E_7$ then split as
\begin{itemize}
\item 36 with $\alpha(R)=0$ to contribute towards building
  $\frak{e}_6$.
\item the $\mathbf{27}$ of $\frak{e}_6$ with $\alpha(R)=2$.
\end{itemize}

\begin{table}
  \begin{tabular}{|c|c|c|c|c|}
    \hline
    &&$R$&$(R,R)$&$\alpha(R)$\\
    \hline
    M-theory ($\alpha_2$) & $E_8$ & $H_1+H_2+H_3+H_4+H_5+H_6+H_7+5H_8$&32
    & 2\\
    & $E_7$ & $H_1+H_2+H_3+H_4+H_5+H_6-2H_7+2H_8$ & 14 & 2\\
    & $E_6$ & $H_1+H_2+H_3+H_4+H_5-H_6-H_7+H_8$ & 8 & 2\\
    & $D_5$ & $2H_1+2H_2+2H_3+2H_4-H_5-H_6-H_7+H_8$ & 20 & 4\\
    & $A_4$ & $5H_1+5H_2+5H_3-H_4-H_5-H_6-H_7+H_8$ & 80 & 10\\
    & $A_1+A_2$ & $H_1+H_2$ & 2 & 2\\
    \hline
    String ($\alpha_1$) & $E_8$ & $H_8$ & 1 & $\ff12$ \\
    & $E_7$ & $-H_7+H_8$ & 2 & $1$ \\
    & $E_6$ & $-H_6-H_7+H_8$ & 3 & $\ff32$ \\
    & $D_5$ & $-H_5-H_6-H_7+H_8$ & 4 & $2$ \\
    & $A_4$ & $-H_4-H_5-H_6-H_7+H_8$ & 5 & $\ff52$ \\
    & $A_1+A_2$ & $-H_3-H_4-H_5-H_6-H_7+H_8$ &6 & $3$ \\
    \hline
    Circle ($\alpha_{\textrm{max}}$) & $E_8$ & $H_7+H_8$ & 2 & 1 \\
    & $E_7$ & $2H_6-H_7+H_8$ & 6 & 2 \\
    & $E_6$ & $3H_5-H_6-H_7+H_8$ & 12 & 3 \\
    & $D_5$ & $4H_4-H_5-H_6-H_7+H_8$ & 20 & 4 \\
    & $A_4$ & $5H_3-H_4-H_5-H_6-H_7+H_8$ & 30 & 5 \\
    & $A_1+A_2$ & $6H_2-H_3-H_4-H_5-H_6-H_7+H_8$ & 42 & 6 \\
    \hline
  \end{tabular}
  \caption{Parabolic Reductions of $E_{n+1}$}  \label{tab:En}
\end{table}

In table \ref{tab:En} we list all the possibilities. Note that we only
list the minimal positive value of $\alpha(R)$. For larger $n$'s we
can get spurious representations appearing as in the $E_7$ example. Note
also that $R$ is only defined up to a multiple. What is well-defined
is $\alpha(R)^2/(R,R)$. We see the pattern
\begin{equation}
  \frac{\alpha(R)^2}{(R,R)} = \frac{8-n}{n+1}, \frac{8-n}{4},
  \frac{8-n}{9-n},
\end{equation}
for $E_{n+1}$ when removing $\alpha_2$, $\alpha_1$ and
$\alpha_{\textrm{max}}$ respectively.

%\bibliographystyle{my-phys}
%\bibliography{string0,new}

\begin{thebibliography}{10}

\bibitem{Sez:survey}
E.~Sezgin,
\newblock {\em Survey of Supergravities},
\newblock arXiv:2312.06754.

\bibitem{Jul:dis}
B.~Julia,
\newblock {\em Group Disintegrations},
\newblock in S.~Hawking and M.~Ro{\v c}ek, editors, ``Superspace and
  Supergravity'', pages 331--350, Cambridge University Press, 1981.

\bibitem{Borel:decomp}
A.~Borel,
\newblock {\em Some Metric Properties of Arithmetic Quotients of Symmetric
  Spaces and an Extention Theorem},
\newblock J. Diff. Geom. {\bf 6} (1972) 543--560.

\bibitem{AM:K3p}
P.~S. Aspinwall and D.~R. Morrison,
\newblock {\em String Theory on K3 Surfaces},
\newblock in B.~Greene and S.-T. Yau, editors, ``Mirror Symmetry II'', pages
  703--716, International Press, 1996,
\newblock hep-th/9404151.

\bibitem{me:lK3}
P.~S. Aspinwall,
\newblock {\em K3 Surfaces and String Duality},
\newblock in C.~Efthimiou and B.~Greene, editors, ``Fields, Strings and
  Duality, TASI 1996'', pages 421--540, World Scientific, 1997,
\newblock hep-th/9611137.

\bibitem{Dabholkar:1998kv}
A.~Dabholkar and J.~A. Harvey,
\newblock {\em {String Islands}},
\newblock JHEP {\bf 02} (1999) 006, hep-th/9809122.

\bibitem{Knapp:beyond}
A.~W. Knapp,
\newblock {\em Lie Groups Beyond an Introduction}, Progress in Mathematics~{\bf
  140},
\newblock Birkh{\"a}user, 2002.

\bibitem{Helgason:Lie}
S.~Helgason,
\newblock {\em Differential Geometry, Lie Groups, and Symmetric Spaces},
\newblock Academic Press, 1978.

\bibitem{CHS:metric}
P.~Candelas, T.~H{\"u}bsch, and R.~Schimmrigk,
\newblock {\em Relation between the Weil-Petersson and Zamolodchikov Metrics},
\newblock Nucl. Phys. {\bf B329} (1990) 583--590.

\bibitem{Tod:inv}
A.~Todorov,
\newblock {\em Applications of K{\"a}hler--Einstein--{C}alabi--{Y}au Metric to
  Moduli of K3 Surfaces},
\newblock Inv. Math. {\bf 61} (1980) 251--265.

\bibitem{Mor:Katata}
D.~R. Morrison,
\newblock {\em Some Remarks on the Moduli of {K3} Surfaces},
\newblock in K.~Ueno, editor, ``Classification of Algebraic and Analytic
  Manifolds'', Progress in Math.~{\bf 39}, pages 303--332, Birkh{\"a}user,
  1983.

\bibitem{Kobayashi-Todorov}
R.~Kobayashi and A.~N. Todorov,
\newblock {\em Polarized Period Map for Generalized {K3} Surfaces and the
  Moduli of {E}instein Metrics},
\newblock T{\^o}hoku Math. J. (2) {\bf 39} (1987) 341--363.

\bibitem{Cand:mir}
P.~Candelas and X.~C. de~la Ossa,
\newblock {\em Moduli Space of {C}alabi--{Y}au Manifolds},
\newblock Nucl. Phys. {\bf B355} (1991) 455--481.

\bibitem{Ramgoolam:1998vc}
S.~Ramgoolam and D.~Waldram,
\newblock {\em {Zero Branes on a Compact Orbifold}},
\newblock JHEP {\bf 07} (1998) 009, hep-th/9805191.

\bibitem{CJ:SO(8)}
E.~Cremmer and B.~Julia,
\newblock {\em The $SO(8)$ Supergravity},
\newblock Nucl. Phys. {\bf B159} (1979) 141--212.

\bibitem{N:torus}
K.~S. Narain,
\newblock {\em New Heterotic String Theories in Uncompactified Dimensions $<$
  10},
\newblock Phys. Lett. {\bf 169B} (1986) 41--46.

\bibitem{Moore:ZVol}
G.~W. Moore,
\newblock {\em Computation Of Some Zamolodchikov Volumes, With An Application},
\newblock arXiv:1508.05612.

\bibitem{Shenker:1990uf}
S.~H. Shenker,
\newblock {\em The Strength of Nonperturbative Effects in String Theory},
\newblock in ``Cargese Study Institute: Random Surfaces, Quantum Gravity and
  Strings'', pages 809--819, 1990.

\bibitem{Kreuzer:1993uy}
M.~Kreuzer and H.~Skarke,
\newblock {\em {ADE String Vacua with Discrete Torsion}},
\newblock Phys. Lett. B {\bf 318} (1993) 305--314, hep-th/9307145.

\bibitem{Eg:K3}
T.~Eguchi, H.~Ooguri, A.~Taormina, and S.-K. Yang,
\newblock {\em Superconformal Algebras and String Compactification on Manifolds
  with $SU(n)$ Holonomy},
\newblock Nucl. Phys. {\bf B315} (1989) 193--221.

\bibitem{Marcus:1983hb}
N.~Marcus and J.~H. Schwarz,
\newblock {\em {Three-Dimensional Supergravity Theories}},
\newblock Nucl. Phys. B {\bf 228} (1983) 145.

\bibitem{mndog:triples}
J.~de~Boer, R.~Dijkgraaf, K.~Hori, A.~Keurentjes, J.~Morgan, D.~R. Morrison,
  and S.~Sethi,
\newblock {\em Triples, Fluxes, and Strings},
\newblock Adv. Theor. Math. Phys. {\bf 4} (2002) 995--1186, hep-th/0103170.

\bibitem{Sei:K3}
N.~Seiberg,
\newblock {\em Observations on the Moduli Space of Superconformal Field
  Theories},
\newblock Nucl. Phys. {\bf B303} (1988) 286--304.

\bibitem{CHL:bigN}
S.~Chaudhuri, G.~Hockney, and J.~D. Lykken,
\newblock {\em Maximally Supersymmetric String Theories in $D<10$},
\newblock Phys. Rev. Lett. {\bf 75} (1995) 2264--2267, hep-th/9505054.

\bibitem{CP:ao}
S.~Chaudhuri and J.~Polchinski,
\newblock {\em Moduli Space of CHL Strings},
\newblock Phys. Rev. {\bf D52} (1995) 7168--7173, hep-th/9506048.

\end{thebibliography}

\end{document}